\documentclass{aa}
\usepackage{natbib}
\usepackage{graphicx}
\usepackage{longtable}
\usepackage{txfonts}
\bibpunct[;]{(}{)}{;}{a}{}{;}

\begin{document}

\title{Supernova rates from the Southern inTermediate Redshift ESO Supernova Search (STRESS)
	\thanks{Based on observations collected at the European
	Southern Observatory, using the 2.2m $MPG/ESO$ telescope on the La Silla (ESO Programmes 62.H-0833,
	63.H-0322, 64.H-0390, 67.D-0422, 68.D-0273, 69.D-0453, 72.D-0670, 72.D-0745, 73.D-0670, 74.A-9008, 75.D-0662) and using Very Large Telescope on the Cerro Paranal (ESO Programme 74.D-0714).}}

\author{M.T.~Botticella\inst{1,2,6}, M.~Riello\inst{3}, E.~Cappellaro\inst{4},
S.~Benetti\inst{4}, G.~Altavilla\inst{5}, A.~Pastorello\inst{6}, M.~Turatto\inst{4}, L.~Greggio\inst{4}, F.~Patat\inst{7}, S.~Valenti\inst{7,8}, L.~Zampieri\inst{4}, A.~Harutyunyan\inst{4}, G.~Pignata\inst{9}, S.~Taubenberger\inst{10}} 

\offprints{M.T. Botticella, \email{botticella@oa-teramo.inaf.it}}

\institute{INAF - Osservatorio Astronomico di Collurania-Teramo,
  V.M. Maggini, I-64100, Teramo, Italy,
\and  Dipartimento di Scienze della Comunicazione, Universit\'a di Teramo, 
  viale Crucioli 122, I-64100 Teramo, Italy
\and Institute of Astronomy, Madingley Road, Cambridge, UK 
\and  INAF - Osservatorio Astronomico di Padova, Vicolo dell'Osservatorio, 5, I-35122, Padova, Italy
\and INAF - Osservatorio Astronomico di Bologna, V. Ranzani 1, I-40127, Bologna, Italy
\and Astrophysics Research Centre, School of Mathematics and Physics, Queen's University Belfast, Belfast BT 71NN, United Kingdom
\and European Southern Observatory, K. Schwarzschild Str. 2,  85748 Garching, Germany
\and  Dipartimento di Fisica - Universit\'a di Ferrara, via del Paradiso 12,  I-44100 Ferrara, Italy
\and Departamento de Astronomia, Universidad de Chile, Casilla 36-D, Santiago,
Chile
\and 
Max-Planck-Institut f{\"u}r Astrophysik, Karl-Schwarzschild-Str. 7, D-85741, Garching bei M{\"u}nchen, Germany
}

\date{Received .../ Accepted ...}

\abstract{}
{To measure the supernova (SN) rates at intermediate redshift we performed the Southern inTermediate Redshift ESO Supernova Search (STRESS). Unlike most of the current high redshift SN searches, this survey was specifically designed to estimate the rate for both type Ia and core collapse (CC) SNe.} 
{We counted the SNe discovered in a selected galaxy sample measuring SN rate per unit blue band luminosity.
Our analysis is based on a sample of $\sim 43000$ galaxies and on 25 spectroscopically confirmed SNe plus 64 selected SN candidates. Our approach is aimed at obtaining a direct comparison of the high redshift and local rates and at investigating  the dependence of the rates on specific galaxy properties, most notably their colour.}
{The  type Ia SN rate, at mean redshift $z=0.3$, amounts to 0.22$^{+0.10 +0.16}_{-0.08 -0.14}$ $h_{70}^2$ SNu, while the CC SN rate, at $z=0.21$, is 0.82$^{+0.31 +0.30}_{-0.24 -0.26}$ $h_{70}^2$ SNu. The quoted errors are the statistical and systematic uncertainties.  } 
{With respect to local value, the CC SN rate at $z=0.2$ is higher
by a factor of $\sim 2$ already at redshift , whereas the type Ia SN rate remains almost constant.  This implies that a significant fraction of SN~Ia progenitors has a lifetime longer than $2-3$ Gyr.
We also measured the SN rates in the red and blue galaxies and found that the SN~Ia rate seems to be constant in galaxies of different colour, whereas the CC SN rate seems to peak in blue galaxies, as in the local Universe.

SN rates per unit volume were found to be consistent with other measurements showing a steeper evolution with redshift for CC SNe with respect to SNe~Ia.

Finally we have exploited the link between SFH and SN rates to predict the evolutionary behaviour of the SN rates and compare it with the path indicated by observations.
We conclude that in order to constrain the mass range of CC SN progenitors and SN Ia progenitor models it is necessary to reduce the uncertainties in the cosmic SFH. 
In addition it is important to apply a consistent dust extinction correction both to SF and to CC SN rate and to measure SN~Ia rate in star forming and in passive evolving galaxies in a wide redshift range.
}


\titlerunning{STRESS}
\authorrunning{Botticella et al.}
\maketitle

\section{Introduction}

SNe, the catastrophic explosions that terminate the life of some stars, play a pivotal role in several astrophysical topics. They provide a crucial test for stellar evolution theory and, as the main contributors of heavy elements in the Universe, they are key actors of the chemical enrichment of galaxies. Their rapidly expanding ejecta sweep, compress and heat the interstellar medium causing gas outflows from galaxies and triggering the star formation process.
Furthermore, in the last few years type Ia SNe have consolidated their prominence as cosmological probes providing the first direct evidence for the acceleration of the Universe \citep{Perlmutter99,Riess98}.  

SN statistics is another important cosmological probe even though less exploited.  
In particular the SN rate as a function of the cosmic time is linked to some of the basic ingredients of the galaxy evolution such as mass, SFH, metallicity and environment.

The rate of CC SNe, including type II and type Ib/c, gives, for an assumed initial mass function (IMF), a direct measurement of the on-going SFR because of the short evolutionary lifetimes of their progenitors. On the other hand the rate of SNe~Ia echoes the long term SFH because these SNe, originating from low mass stars in binary systems, are characterized by a wide range of delay times between progenitor formation and explosion \citep{Yungelson,Greggio05}.

Conversely, if a SFH is assumed, the progenitor scenarios for SN~Ia can be constrained measuring their rate as a function of galaxy type and redshift \citep{Madau98,Dahlen99,Pain02,Strolger,Maoz,Tonry,Scannapieco,Mannucci05,Mannucci06,Neill06,Sullivan06,Forster}.
Despite promising prospects, measurements of SN rates were very scanty so far.

A new opportunity for SN searches emerged in recent years thanks to the availability of panoramic CCD mosaic cameras mounted at medium/large size telescopes. These instruments allow the deep monitoring of wide areas in the sky and thus the collection of statistically significant samples of SNe at intermediate and high redshifts.   

In this paper we report the results of the "Southern inTermediate Redshift ESO Supernova Search" (STRESS) which was carried out with ESO telescopes and designed to search for SNe in the range $0.05<z<0.6$.

The paper is organized as follows: in Sect.~2 we illustrate the aims and the strategy of STRESS, in Sect.~3 we present the data set collected during our observing programme. The following three sections describe the data analysis, namely the selection of the galaxy sample (Sect.~4), the detection and classification of the SN candidates (Sect.~5) and the estimate of the  control time of the search (Sect.~6). The measurement of the SN rates and the estimate of the statistical and systematic uncertainties are presented in Sect.~7. Our results are compared
with published measurements and models predictions in Sect.~8. Finally, a brief summary is given in Sect.~9.

We remark that the measurements reported in this paper supersede the preliminary results published in \citet{Cappellaro05} because of $i)$ better SN statistics ,  $ii)$ better filter coverage for characterization of the galaxy sample, $iii)$ improved correction for the host galaxy extinction. 

Thorough the paper we adopt the cosmological parameters $H_0\footnote{$h_{70}=H_0/70 \,{\rm km}\,{\rm s}^{-1} {\rm Mpc}^{-1}$}=70\,{\rm km}\,{\rm s}^{-1} {\rm Mpc}^{-1}$, $\Omega_M=0.3$, $\Omega_\Lambda=0.7$. Magnitudes are in the Vega System.

\section{Search goals and strategy}\label{stress}

Most past and current high-redshift SN searches were/are aimed at detecting SNe~Ia for measuring cosmological parameters, while the measurement of SN rates is a secondary goal. As such, the strategy  of these searches are affected by several shortcomings \citep[eg. ][]{Schmidt,Pain02,Tonry,Dahlen04,Neill06}. In particular, high-redshift SN searches are optimized for the detection of un-reddened SNe~Ia near maximum light, so that all variable sources unlikely to fall into this category are assigned a lower priority for follow-up, when not completely neglected. This strategy introduces a severe bias in the SN sample; besides, in general, it prevents a reliable estimate of the rate of CC SNe.

STRESS was specifically designed to measure the rate of both SNe~Ia and CC SNe at intermediate redshift, to be compared with the local values and estimate their evolution. To this aim we tried to reduce as much as possible the biases which affect the different SN types.

To measure the SN rate we counted the events observed in a selected galaxy sample. This is the same approach used to derive the local rates, and offers two advantages.
By preserving the link between the observed SNe and the monitored galaxies we can investigate the dependence of the rates on specific galaxy properties, most notably their colours, which depend primarily on the galaxy SFH, and on metallicity and dust extinction.
In addition, the SN rates measured per unit of galaxy $B$ luminosity, that is in SN unit (SNu\footnote{ 
$ SNu = SN\, (100 {\rm yr})^{-1} \,(10^{10}{\rm L}_\odot^{\rm B})^{-1}$ }), can be directly compared with the local ones.

For the selection and the characterization of the galaxy sample, we used multi-band observations and the photometric redshift technique ({\em photo-z}), which allow us to derive the redshift, the absolute luminosity, and the rest frame colours for all the galaxies in our sample.

In summary our approach 
involves the following steps: 
\begin{enumerate}
\item the selection of the galaxy sample and its characterization,
\item the detection and classification of SN candidates,
\item the estimate of the control time of the galaxy sample,
\item the measurement of SN rates, the analysis of their dependence on the galaxy colours, and their evolution with redshift.
\end{enumerate}

\section{The observing programme}\label{obs}

Observations were carried out using the Wide Field Imager (WFI) at the 2.2m MPG/ESO telescope at La Silla, Chile. WFI is a mosaic camera consisting of $2\times 4$ CCDs, each of $2048\times 4096$ pixels, with a pixel scale of 0.238 arcsec and a field of view of $34\times 33$ $arcmin^2$ . The individual chips are separated by gaps of 23.8 arcsec and 14.3 arcsec along right ascension and declination respectively, for a resulting filling factor of $95.9\%$. 

We performed observations in the $B,V,R,I$ bands using the following ESO/WFI broad-band filters: $B/99$, $B/123$, $V/89$, $Rc/162$, $Ic/lwp$.

Our initial list included 21 sky-fields evenly distributed in right ascension to reduce the observing scheduling requirements. The fields were chosen at high galactic latitude to reduce stellar crowding and galactic extinction and with few bright sources to minimize CCD saturation and ghost effects.

The observing programme, distributed over a period of 6 years, from 1999 to 2005, can be divided into three phases. During the first year we carried out a pilot programme aimed at tuning the observing strategy and testing our software (at the beginning of our programme not all filters were available). In a second phase (2001-2003, 12 observing runs) we performed the SN search in the $V$ band targeting events at redshift $z\sim 0.2-0.3$. In the last phase (2004-2005, 4 observing runs), we performed the search  in  $R$ band aiming to detect SNe at higher redshifts. A total of 34 nights were allocated to our programme out of which 9 were lost due to bad weather. Seeing was in the range $0.7-1.5$ arcsec, with an average around 1 arcsec.

\begin{table}
\caption{List of the fields with multi-band coverage.}
\label{tabfields}
\centering
\begin{tabular}{l c c c l}
\hline\hline
Name & R.A. (J2000.0) & Dec. (J2000.0) & E(B-V)$^\#$ & Band \\
\hline
03Z3 &03:39:31.1 & $-$00:07:13 &0.081   & $B,V,R,I$\\
10Z2 &10:46:45.8  & $-$00:10:03 &0.040  & $B,V,R$\\
13Z3 &13:44:28.3  & $-$00:07:47 &0.026  & $B,V,R$\\
22Z3 &22:05:05.8  & $-$18:34:33 &0.026  & $B,V,R$\\
AXAF &03:32:23.7 & $-$27:55:52 &0.008   & $B,V,R,I$\\
EF0427 &04:29:10.7 &$-$36:18:11 &0.022  & $B,V,R,I$\\
EisA1 &22:43:25.6 &$-$40:09:11 & 0.014  & $B,V,R$\\
EisB &00:45:25.5  & $-$29:36:47 &0.018  & $B,V,R$\\
EisC &05:38:19.8 & $-$23:50:11& 0.030   & $B,V,R,I$\\
EisD & 09:51:31.4 & $-$20:59:25 & 0.040 & $B,V,R,I$\\
Field2 & 19:12:51.9 & $-$64:16:31 &0.037& $B,V,R$\\
J1888 & 00:57:35.4 & $-$27:39:16 &0.021 & $B,V,R$\\
white23 &13:52:56.4 &$-$11:37:03 &0.067 & $B,V,R,I$\\
white27 &14:10:59.7& $-$11:47:35 &0.057 & $B,V,R,I$\\
white31 &14:20:15.3 &$-$12:35:43 &0.096 & $B,V,R,I$\\
whitehz2 &13:54:06.1& $-$12:30:03&0.073 & $B,V,R,I$\\
\hline
\end{tabular}

$\#$  Galactic extinction from \citet{Schlegel}
\end{table}

\addtocounter{table}{1}
The temporal sampling, on average one observation every 3-4 months, was tuned to maximize event statistics. The observing log is reported in Tab~\ref{logobs}.

\onllongtab{2}{
\begin{longtable}{lccccccl}
\caption{\label{logobs}Observing log of the SN search fields.} \\
\hline\hline
 run  & Field & Filter & Nexp  & Texp  & Seeing  \\
\hline
\endfirsthead
\caption{continued.}\\
\hline\hline
 run  & Field & Filter & Nexp  & Texp  & Seeing  \\
\hline
\endhead
\hline
\endfoot
1999-02-23 &  10Z2 & B & 3 & 1200 & 1.12\\
& 13Z3  & B & 3 &1200  & 1.10\\
1999-03-10 & 10Z2 & B  &3  & 1200 &1.14 \\
& 13Z3 & B & 3 &1200   & 1.15\\
& 10Z2 &V  &3  & 1200 &1.25 \\
&13Z3 & V  & 3 & 1200 & 1.15\\
1999-03-19 & 10Z2 & B & 3 &1200  &1.04 \\
& 13Z3 &B  &3  &1200  & 0.96\\
& 10Z2 &V & 3 & 1200  & 0.86  \\
& 13Z3 & V &3  & 1200 &0.84 \\
1999-05-08 & 10Z2 &B  &  3 & 1200 & 1.34 \\
&13Z3 & B  & 3  & 1200 & 1.4 \\
& Field2   &  B &3  & 1200 &1.91 \\
& Field2   & R  &3  & 1200  & 1.02  \\
& 10Z2 & V & 3 &  1200 & 1.37\\
& Field2 &V  &3  &1200  &1.62 \\
1999-05-17 & 10Z2 & B  &3  &1200  &0.81 \\
& 13Z3 &  B & 3 &1200  & 1.74 \\
& Field2   &B  &3  &1200  & 0.74\\
& 10Z2 & R  & 3  &1200  & 0.96 \\
& 13Z3 & R &3  &1200   & 0.91\\
& 10Z2 & V & 3  &1200  &0.90 \\
& 13Z3 & V & 3 &1200  &0.98 \\
& Field2 &V  & 3  &1200  &0.81 \\
1999-08-03 & 13Z3 &  B &3  & 1200 & 1.43\\
& 22Z3 & B  & 3  & 1200 &1.17 \\
& Field2 & B  &3  &1200  &1.12 \\
&J1888 & B  &3  &720 & 1.22 \\
&J1888 & R & 3 &720  & 1.31\\
&Field2 &V  &  3 &1200  &1.8 \\
&J1888 &  V & 3 & 720  &1.41 \\
1999-08-14 & 22Z3 & B & 2   &1200  &1.58 \\
& EisA1 &  B  & 3 &1200  &1.67 \\
&Field2  & B  &3  & 1200 & 2.15  \\
1999-08-31& 22Z3 & B  &3  & 1200 & 1.43\\
& Field2 & B &3  & 1200  &2.44 \\
1999-09-13 & J1888 & B  & 5  &720  & 1.62\\
&J1888 & R &  5 &  720 &1.24 \\
&22Z3 &V  &3  & 1200  & 1.46 \\
& EisA1 & V  & 3 &1200  &1.32 \\
& EisB & V  &3  &1200  &1.27 \\
&Field2 & V & 3  & 1200 &1.67 \\
&J1888 & V & 5  & 720   & 1.41 \\
1999-10-30& AXAF & B &3  & 900 &3.00 \\
&EisA1 & B  & 3 & 900 & 1.32 \\
& EisB &  B &3  & 900 &2.02 \\
&EisA1 & V & 3 & 900  & 1.50\\
&EisB & V &3  & 900 &1.52 \\
1999-11-09& 03Z3 & B & 3 &900  &0.98 \\
&AXAF &B   &3  &900  &1.13 \\
&EisA1 & B  &3  &900  & 1.12\\
&J1888 &B  &3  &900  &  1.05\\
&03Z3 &V  &3  &900  &1.07 \\
&AXAF &V  & 3  &900  &1.08 \\
&EisA1 & V & 3  &900  &0.84 \\
& EisB & V   & 3 & 900 & 0.76\\
& J1888 & V & 3   &900  &0.81 \\
1999-12-02& AXAF & B  &3  &900  &1.03 \\
& AXAF &  V & 3  &  900 & 1.12\\
&EisB &V  & 3  &900  &1.27 \\
& EisC &V  &3   & 900  &1.15 \\
&EisD & V  &3  & 900  & 1.2\\
&J1888 & V   &  3 &900   & 1.06 \\
1999-12-10& 03Z3 & B & 3  & 900 & 1.6\\
&03Z3 & V  &3   &900  &1.62 \\
&10Z2 & V & 3 &900   &1.28 \\
&EisA1 &V  &3  &900  &1.75 \\
& EisC &V  &3   & 900  &1.31 \\
& J1888 &  V & 3  &900  &1.65 \\
1999-12-28&EisC & B & 3  &900  &1.18  \\
& 03Z3& V &  3 &1200  & 2.18 \\
& AXAF &V  &3  &900&  1.15 \\
&EisC & V  &3  &900  &0.77 \\
&EisD & V &3   &900  &0.96 \\
&J1888 & V  & 3 &900  &1.10 \\
2000-01-08 & EisC & B  &3  & 900 &1.06 \\
&EisD & B  & 3  &900  & 0.91 \\
&EisC &V   & 3 &900  & 0.85\\
&EisD &V  &3  &900  & 0.90\\
&10Z2 &V &  3 & 900 & 0.92  \\
2000-11-16& 03Z3 & V & 3 &900&  1.14 \\
&AXAF &V  & 3  & 600 &0.92 \\
& EF0427 & V & 3  & 900  &0.74 \\
&EisA1 & V & 3 &900  &0.92 \\
&EisB & V & 3  &900  &0.86 \\
&EisC &V  & 3  &900  &0.64 \\
&EisD &  V &3  &900  & 0.83\\
&J1888 &  V &3  &900  & 1.05 \\
2000-12-17& 03Z3 &V  &3  &600  &1.14 \\
&AXAF & V &  3 & 600 &0.93 \\
&EF0427 &V  & 3 & 900 &0.93 \\
&EisA1& V &3  &900  & 1.12\\
&EisB & V  &  3 &900  &1.26 \\
&EisC & V &  3&  900& 0.85  \\
&J1888 & V  & 3 & 660 &1.02 \\
2001-04-18& 10Z2 & V & 3 & 900 & 0.98 \\
&13Z3 &  V &  3  &900  & 0.85\\
&22Z3 & V &  3 & 900 &1.41 \\
&EisC & V & 3  & 900 &0.85 \\
& EisD &  V &3  & 900  &0.86 \\
&Field2 & V   & 3 & 900  &0.84 \\
& white23 &   V & 3 &900 & 0.82\\
2001-04-19& white27 &  R & 3 &1200  & 0.71  \\
& whitehz2 &R  & 3 &1200  & 0.67\\
& white27 & V &3  &900  &0.73 \\
& white31 & V  & 3 & 900 & 1.01\\
& whitehz2 & V & 3 & 900 &0.79 \\
2001-11-11& 03Z3 & V  & 3 & 900  & 0.90 \\
& 22Z3 & V   & 3  & 900 & 0.84\\
& EF0427 & V &3  &900  & 0.75\\
& EisA1 & V  &3   &900  &0.82 \\
&EisB & V & 3 & 900  &0.78 \\
&EisC & V  & 3  & 900  &0.91  \\
&J1888 &V  &3  &900  &0.75 \\
2001-11-12& 22Z3 &  R &3  &900  &1.31 \\
&AXAF & R & 3  &900  &0.92 \\
&EisA1 &R & 3  & 900 & 1.55  \\
&EisB &  R & 3 & 900 &1.07 \\
&J1888 & R & 3 & 900 & 1.24\\
&AXAF &  V & 3 &900  &1.03 \\
2001-11-18& 03Z3 &V &3 & 900   & 0.90  \\
&22Z3 &V  & 3  &900  & 1.10\\
& AXAF &V  & 3 & 900 & 0.84\\
& EF0427 & V  & 3  & 900 &0.93 \\
& EisA1 & V &  3 &  900 &1.20 \\
&EisB &V  & 3 &900  & 0.83  \\
&EisC & V & 3 &900   &0.65 \\
&J1888 & V &  3 &900  &1.02 \\
2001-12-08&03Z3 & V  & 3 &900  &0.94 \\
& AXAF &V  & 3  &900   &0.95 \\
& EF0427 &  V & 3 & 900 &0.95 \\
&EisB & V &3  & 900 & 0.88 \\
&EisC & V &3  & 900 &  0.95\\
& J1888 &  V &  3  & 900 & 0.94\\
2001-12-09& 22Z3 & R &1  &900  &  2.00\\
& AXAF&  R & 3  & 900 &0.82 \\
& EF0427 & R   & 3 & 900 & 1.1\\
&EisB & R & 3 & 900 &0.89 \\
&EisC &  R &  3 & 600  & 1.05\\
&J1888 &  R &3  & 900 &0.84 \\
& 22Z3 & V & 1 &900  & 1.90\\
&EisD & V  & 3 & 900 &0.92 \\
2002-04-07&EisC & V  &3  & 900 &1.32 \\
&EisD & V &3  & 900  & 0.92\\
&10Z2 & V & 3 & 900 &0.91 \\
&13Z3 & V  & 3 & 900 &0.78 \\
&white27 & V &3  & 900  &0.80 \\
& white31 & V  & 3 &900  &0.85 \\
&Field2 &  V & 3  & 900 &1.30  \\
2002-04-08& EisD & R  &  3 & 900  &0.85\\
& 10Z2 & R & 3 & 900 & 0.88  \\
& 13Z3 & R & 3 &900  &1.1 \\
&white31 & R  & 3 &900  &0.87 \\
&Field2 & R &  3 & 900 &1.31 \\
&white23 &V  & 3 &900  & 0.96\\
&whitehz2 &V  &3  &900  &0.91 \\
2004-02-18  & EF0427 & R &3 &900  &0.85 \\
& EisC & R &3 &900 &0.91 \\
& EisD &R &3 &900 &0.93  \\
 & white27 & R &3 &900 &0.82  \\
& white31 &R &3 &900 &0.94  \\
 & white31&R &3 & 900&0.91  \\
  & whitehz2 &R  & 3 &900 &0.83  \\
& 10Z2 & R &  3& 900&0.87  \\
2004-02-19  &EF0427 & I&3 &900 &1.05 \\
& EisC & I &3 & 900& 0.80 \\
 &EisD &I &3 &900 & 1.03  \\
  &10Z2 & I & 3&900 &0.73 \\
  &whitehz2 & I  &3 &900 &0.76 \\
  &white27 & I& 3 &900 & 0.67 \\
  &white31 & I &3 &900 & 0.75\\ 
 2004-05-14 & 13Z3&R &3 &900 &0.78 \\  
  &EisD & R &3 &900 &1.11 \\ 
  & 22Z3 & R & 3 &900 &0.86 \\ 
  & white31 &  R&3 &900 &1.18 \\ 
  &whitehz2 & R& 3&900 &0.78 \\
  &white27 &R  &3 &900 &0.81 \\
 2004-05-15 & EisD & B &3 &900 &1.57 \\
  &white31 &B & 3 &900  &1.38 \\
  &whitehz2 &B & 3&900 &1.38 \\ 
  &white27 & B &3 &900  &1.55 \\ 
 2004-07-14 &white27 &R &3 &900 & 1.01\\ 
  &white31 &R & 3 & 900&1.12 \\ 
  &22Z3 &R &3 &900& 0.94 \\ 
  &whitehz2 & R &3 & 900 &0.86 \\ 
  &EisA1 & R & 3 & 900 & 0.90\\ 
  &EisB &R & 3& 900&1.02 \\
2005-05-13  &white23 & R&3 & 900 &1.27 \\
  &white27 & R &3 &900 &1.48 \\
  &white31 & R &3 &900 &1.55 \\
  & whitehz2 &R  &3 &900  &1.18 \\
  &13Z3 &R &3 &900 &1.29 \\
  & EisD & R  & 3&900 &1.56 \\
  & 10Z2 &R & 3&900 & 1.39\\
  &Field2 & R& 3&  900&1.32  \\
\end{longtable}
 }

Typically, in each run we observed the same fields for two consecutive nights, in the first one with the {\em search} filter, $V$ or $R$, and in the second one with a different filter to obtain colour information both for SN candidates and galaxies. Each observation, typically of 45~min integration time, was split in a sequence of three exposures jittered by 5-10 arcsec to allow a better removal of cosmetic defects, cosmic rays, satellite tracks and fast moving objects. Due to technical and weather limitations and, occasionally, to scheduling constraints, in many cases we could not maintain our observational strategy. As a consequence, for many SN candidates we are missing colour information and for a few fields we have insufficient filter coverage to apply the {\em photo-z} technique to galaxy sample.

In order to secure the classification of the SN candidates spectroscopic observations were scheduled about one week after the search run. In the first year,
spectra were taken at the ESO3.6m telescope equipped with ESO Faint Object Spectrograph and Camera (EFOSC); in the following years, we used the FOcal Reducer and low dispersion Spectrograph (FORS1/2) at the VLT.
In total 2.5 nights were allocated at the ESO3.6m telescope and 12 nights at the VLT, 4 of which were lost due to bad weather conditions. 

For a better subtraction of the night sky emission lines we selected grisms of moderate resolution (FORS1/2 grism $300V$ and/or $300I$) which allow the sampling of a wide wavelength range (445-865 nm and 600-1100 nm respectively) with a resolution of $\sim10\AA$. The width of the slit was chosen to match the seeing but, to reduce the contamination of the host galaxy background, never exceeded 1 arcsec. Depending on the magnitude of the SN candidate, the exposure time of spectroscopic observations ranges from 900 seconds to 3 hours. 

As we will detail in Sect.~5 we could obtain direct spectroscopic observation only for a fraction ($\sim 40\%$) of the SN candidates. In order to obtain the redshifts of the host galaxies of the remaining SN candidates, and to check for signs of the presence of Active Galactic Nuclei (AGN), we carried out a follow-up program using FORS2. The observing strategy, tuned to make the best use of non-optimal seeing and sky transparency conditions, proved to be very successful: we obtained 44 spectra of SN candidate host galaxies in
a single ESO period.

In addition, a dedicated service program using WFI (12 hours) was designed to complete the photometric band coverage of the fields monitored during SN search. 

Eventually we were able to secure $B,V,R,I$ imaging for 11 fields and $B,V,R$  for 5 fields, whose coordinates, galactic extinction and band coverage are shown in Table~\ref{tabfields}. Further data analysis has been restricted to these 16 fields.  

\section{The galaxy sample}\label{Gsample}

In this section we describe the steps required to select our galaxy sample, namely the detection of galaxies in the monitored fields (Sect.~\ref{Gdetect}), the selection criteria and the application of the {\em photo-z} technique (Sect.~\ref{Gsel}). 
We also present a number of crosschecks with other observations and galaxy catalogues that were performed to validate our catalogues (Sect.~\ref{Cal} and Sect.~\ref{Gsel}). 

\subsection{Source catalogues}\label{Gdetect}

For optimal source detection in each field we obtained deep images in each band.
First we selected all the images with the best seeing and sky transparency for each field and band.  Then we selected, for each band, a seeing range of $ <0.15$ arcsec with the requirement that it contains the maximum number of images which hereafter are co-added.

All images were preliminarily processed using the {\em IRAF}\footnote{IRAF is distributed by the National Optical Astronomy Observatories, which are operated by the Association of Universities for Research in Astronomy, Inc., under cooperative agreement with the
National Science Foundation.} package {\em MSCRED}, specifically designed to handle mosaic CCD images \citep{Valdes}. We followed the standard recipes for CCD data reduction applying bias subtraction, flat field correction, trimming and, for $I$ band images, de-fringing. Astrometric calibration for each image was performed using as reference the USNO--A2 catalog \citep{Monet} with the task {\em msccmatch}. The r.m.s. dispersion of the absolute astrometry is $\sim0.3$ arcsec. 

The photometric calibration was obtained with reference to Landolt standard fields \citep{Landolt} observed on photometric nights. Images obtained under non-photometric conditions were scaled to the flux of calibrated images by matching the photometry of the common stars.

After a proper match of the flux scale, the selected images of a given field and band were stacked together using the {\em SWARP}\footnote{ www.terapix.iap.fr} software \citep{Bertinswarp} with a third-order $Lanczos$ kernel for the resampling, and a median algorithm for the co-adding procedure. 

The limiting magnitudes of the co-added images reflect the very different total exposure times for the search and complementary bands: the 3-$\sigma$ limit for point sources was reached at $B_{lim}\sim23$ mag, $V_{lim}\sim24$ mag, $R_{lim}\sim23$ mag and $I_{lim}\sim22$ mag.  

In order to consistently measure source colours, we have to take into account that the co-added images of a given field show different point spread functions (PSF) in each band. To avoid colour offsets, the co-added images in each band were convolved with an appropriate Gaussian kernel, using the {\em IRAF} task {\em gauss}, to match the PSF of the co-added image with the worst seeing.

For source detection and photometry we used {\em SExtractor} \citep{Bertin} in dual image mode: source detection and classification was obtained from the original $V$ band images, since this is the band with the best $S/N$ ratio, while the magnitudes of the sources were measured on the convolved images using the same adaptive aperture as in the original $V$ images. 
This procedure assures that the fluxes are measured in the same physical region of each source in all bands.  We adopted a source detection threshold of $3 \sigma$ above the background noise and a minimum of 12 connected pixels above the detection threshold.
To measure magnitude we chose a Kron-like elliptical aperture (with 2.5 pixel Kron radius) adaptively scaled to the object dimensions ($mag\_auto$).

The instrumental magnitudes were normalized to 1 sec exposure time and corrected for the instrument zero point, atmospheric and Galactic extinction. 
For our purposes the color term correction in the photometric calibration is not required and, therefore, has not been applied.
 
For each field a general catalogue was created including all detected sources, their magnitude in all bands and the relevant {\em SExtractor} parameters. Close to very bright stars the efficiency and reliability of source detection and photometry are significantly lower than the average. These regions were masked automatically using {\em SExtractor} parameters that describe the shape and the size of the bright stars and sources detected in these areas were deleted from the catalogue. Finally, visual inspection of the detected sources was performed on the images in order to remove spurious detections and to verify the detection uniformity across the effective area.

\begin{figure}
\resizebox{\hsize}{!}{\includegraphics{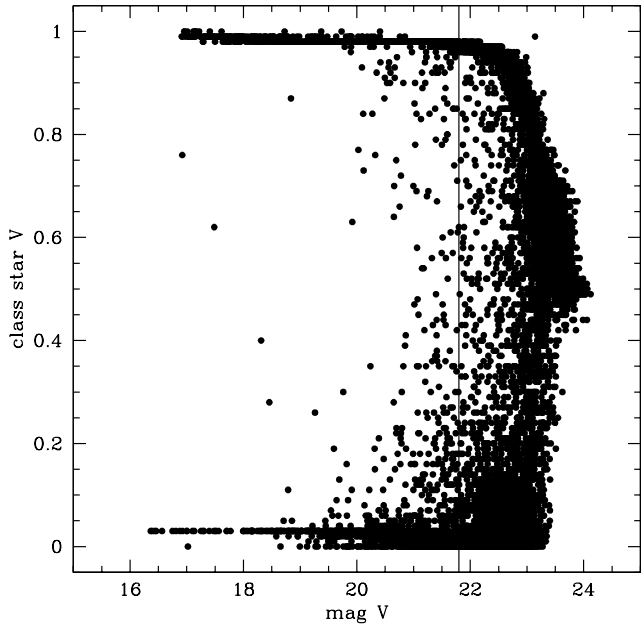}}
\caption{The stellarity index of the sources detected in one of our co-added image as function of magnitude. The vertical line indicates the cut-off magnitude for our galaxy sample.}
\label{classstar}
\end{figure}

To separate stars from galaxies we used the {\em SExtractor} neural-network classifier fed with isophotal areas and the peak intensity of the source.  The classifier returns a parameter named $class\_star$ with a value between 1 (for a point-like object) and 0 (for an extended profile). Fig.~\ref{classstar} shows the $class\_star$ parameter as function of magnitude for the sources detected in one of our co-added image. For bright objects stars are clearly divided by galaxies while for fainter magnitudes the sequences spread out and merge. An intrinsically wide scatter in $class\_star$ is typical of faint galaxies.  

We defined as galaxies all sources having $class\_star< 0.9$ \citep{Arnouts}. In principle this choice may result in an overestimate of the number of galaxies at faint magnitudes but ensures that our sample is not biased against compact galaxies. Notice however that because of the relatively bright cut-off limit adopted for the galaxy sample selection (see Fig.~\ref{classstar} and Sect.~\ref{Gsel}) our analysis is confined to a regime where the star/galaxy separation is reliable.

\subsection{Catalog crosschecks}\label{Cal}

To check the accuracy of the astrometric and photometric calibration of our catalogs we performed a number of comparisons with the results of galaxy surveys and with model predictions.

\begin{figure}
\resizebox{\hsize}{!}{\includegraphics{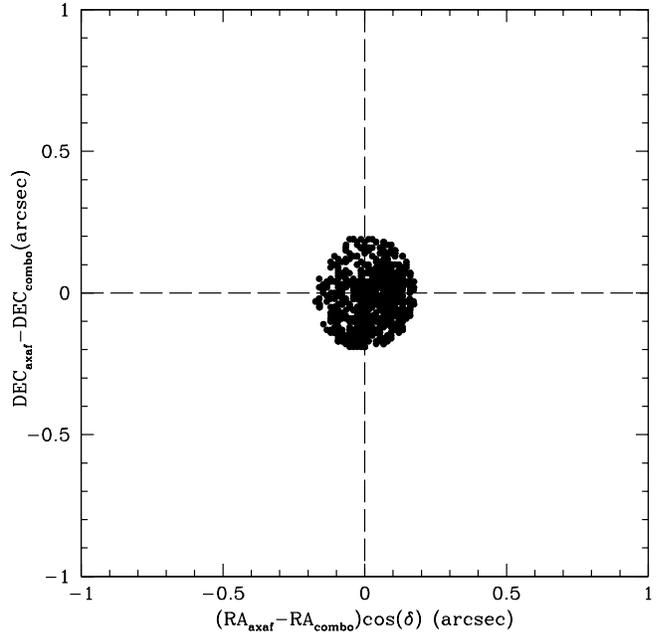}}
\caption{Differences in the equatorial coordinates between our source list for the AXAF field and the COMBO-17 catalog for the objects in common.}
\label{astrcombo}
\end{figure}

\begin{figure}
\resizebox{\hsize}{!}{\includegraphics{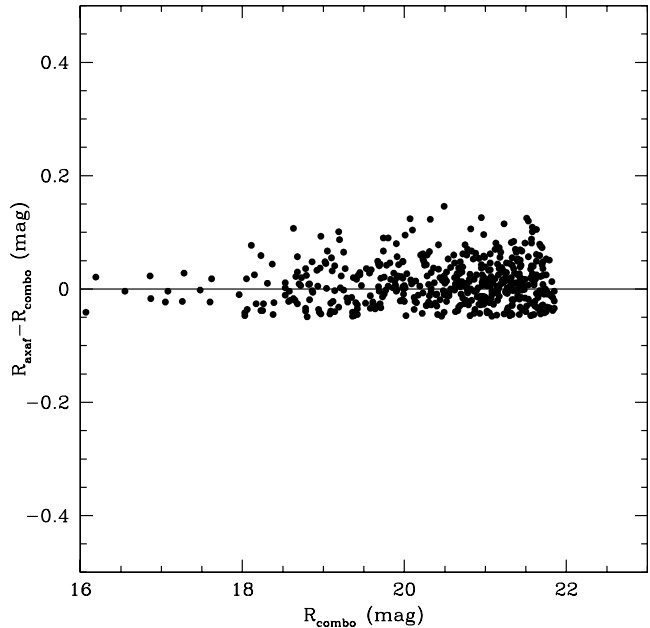}}
\caption{Difference between the R magnitudes of galaxies in our source list for the AXAF field and in the COMBO-17 catalog as a function of the $R_{Combo}$ magnitude for magnitudes brighter than $R_{Combo}=21.8$.}
\label{fotcombo}
\end{figure}

In particular one of our fields, dubbed AXAF, partially overlaps with the Chandra Deep Field South which has bee observed also with WFI at the ESO2.2m telescope during the ESO Imaging Survey \citep[EIS, ][]{Arnouts} and the COMBO-17 survey \citep{Wolf03}.
In the overlapping area we identified $\sim 1200$ galaxies in common with the COMBO-17 $R$ band source list. The comparison of the astrometric calibration between our catalog and COMBO-17 ( Fig.~\ref{astrcombo}) shows a small scatter (r.m.s. $\sim 0.15$ arcsec) both in right ascension and in declination. Also the comparison of the $R$ band photometric calibration (Fig.~\ref{fotcombo}) shows a  small zero point offset ($\sim 0.01$mag) and a small dispersion (rms $\sim 0.03$mag) for magnitudes brighter than the cut-off limit ($R=21.8$ mag).

To verify the photometric calibration for all the other fields, we selected all bright (non saturated) stars, i.e. sources with $class\_star>0.9$ and brighter than $V=19$ mag, and compared them to the Landolt's stars on two colour-colour ($B-V,V-R$,$V-R,R-I$) diagrams. The magnitudes of the Landolt's stars were transformed onto the WFI filter system using the corresponding color terms as reported in the instrument web page\footnote{http://www.ls.eso.org/lasilla/sciops/2p2/E2p2M/WFI/zeropoints}. 
We also checked the observed number counts as a function of colour with those predicted by the galactic star count model of \citet{Girardi} {\em TRILEGAL}\footnote{http://trilegal.ster.kuleuven.be/cgi-bin/trilegal} for the WFI filter's set. Performing these tests for all fields we were able to confirm the photometric calibration to better that $\sim 0.1$mag in all bands.

\subsection{Galaxy sample selection and characterization}\label{Gsel}

Defining a photometric redshift catalogue involves fulfilling two requirements: $i)$ to
photometrically define a sample of galaxies for which reliable photometric redshifts can
be obtained and $ii)$ to characterize the redshift error distribution.

In order to estimate the galaxy redshifts we used the spectral energy distribution (SED) template fitting technique. In practice, a grid of predicted galaxy colours as a function of redshift is constructed performing synthetic photometry on different galaxy spectral templates, either empirical or produced by spectral synthesis modeling.  Then, a $\chi^2$ test of the observed vs. the predicted colors is used to select the best fit redshift and spectral template in the grid for each galaxy of the catalogue. 

The SED fitting procedure is based on the fit of the overall shape of spectra and on the detection of strong spectral features, in particular, the $4000\AA\/$, the Balmer or Lyman continuum breaks. 
Obviously the accuracy of this technique improves with the spectral range coverage: with $B$ to $R$ band photometry reliable redshifts can be obtained only in the range $0.2<z<0.8$, which may be extended to $z\sim1$ by adding the $I$ band photometry.  Redshifts for both very nearby galaxies, which need the $U$ band photometry, and more distant galaxies, which require infrared (IR) filter coverage, are definitely more uncertain.

In order to reduce the contamination from distant galaxies, which may be erroneously placed at low or intermediate redshift, we set a cut-off limit for the galaxy catalogue at an apparent magnitude of $R=21.8$ mag. This corresponds to $M^*_B$ at redshift $z=0.8$, where $M^*_B$ is a parameter of the Schechter function which fits the galaxy luminosity function \citep{Wolf03}. 
We stress that for computating SN rates, we do not need the galaxy catalogue to be complete within a given volume.
Nevertheless we note that at redshift $z\sim 0.2-0.3$, where the
observed SN distribution peaks (see Sect.\ref{classification}), the contribution to the total luminosity from galaxies fainter than $R=21.8$ mag is only $25\%$.    

Photometric redshifts for the selected galaxies were computed using the code {\em hyper-$z$}\footnote{www.web.ast.obs-mip.fr} \citep{Bolzonella} running in the predefined redshift range $0<z<0.8$. 
As spectral templates we tested both the observed spectra of Coleman, Wu $\&$ Weedman (CWW) \citep{Coleman} as well as the synthetic models based on Bruzual $\&$ Charlot (BC) library \citep{Bruzual}. 
Along with the redshift, the code provides the rest frame $B$-band absolute magnitude for the best fitting template.  
To further reduce the contamination from outliers we set a prior on the galaxy absolute luminosity. When $M_B$ is out of the range $(-23,-16)$ mag we examine the second solution provided by {\em hyper-z}, and if this is also inconsistent with the luminosity prior, we remove the galaxy from the catalog, This happened to
$16\%$ of the galaxies in the sample.

Our final galaxy sample counts 43283 objects.

The accuracy of the photometric redshifts is estimated by comparing them
to spectroscopic redshifts, when available, and to the photometric redshifts
 obtained by the COMBO-17 survey, which has a better SED sampling and used a different {\em photo-z} code \citep{Wolf03}. 

\begin{figure}
\resizebox{\hsize}{!}{\includegraphics{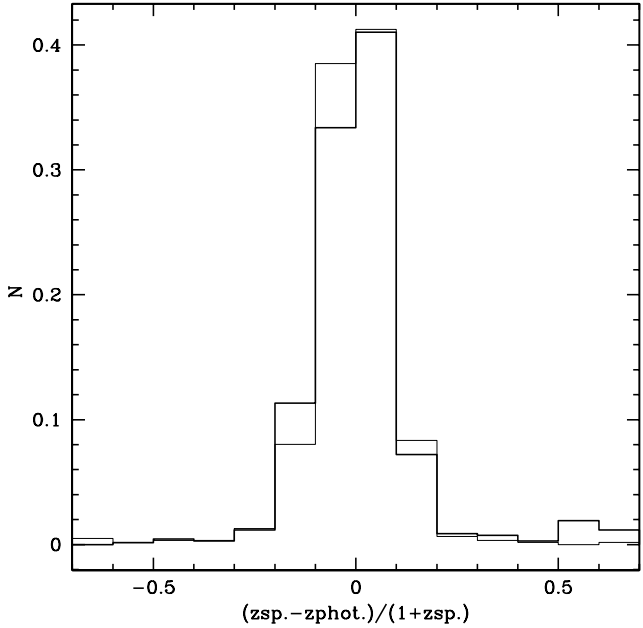}}
\caption{Distribution of the fractional error $\Delta z =(z_{\rm ph}-z)/(1+z)$ in the comparison of our estimates $z_{\rm ph}$ with spectroscopic redshift (thin line) and with photometric redshit of COMBO-17 (thick line).}
\label{sp}
\end{figure}

The spectroscopic redshifts for 470 galaxies of our sample were retrieved from the Nasa/IPAC Extragalactic Database\footnote{http://nedwww.ipac.caltech.edu/}, and complemented with 92 unpublished redshift measurements from the EDISC survey (Poggianti, private communication). We quantify the reliability of the photometric redshifts by measuring the fractional error, $\Delta z =(z_{\rm ph}-z_{\rm sp})/(1+z_{\rm sp})$ where $z_{\rm ph}$ and $z_{\rm sp}$ are the photometric and spectroscopic redshifts, respectively.  The distribution of the fractional error, both in the case of CWW templates (Fig.~\ref{sp}) and in that of BC templates, does not show systematic off-sets: for both cases a
$\langle\Delta z \rangle=0.01$ and  $\sigma(\Delta) \sim 0.12$ appear consistent with our limited SED sampling. This accuracy is sufficient for our statistical analysis. In addition, the fraction of "catastrophic" outliers, galaxies with $\Delta z \ge 3\sigma$, is relatively low, $\sim 3\%$. 

We also tested wether there are systematic trends between the photometric redshift reliability and redshift range and the galaxy spectral type and  magnitude. 
While there is no evidence of systematic trends with the magnitude (because we have selected bright galaxies), the dispersion and the fraction of outliers increase significantly at $z<0.2$, as expected, since we lack the $U$ band coverage.
The dispersion of $\Delta z$ also increases going from early to late type galaxies, due to the  fact that
the {\em photo-z} technique strongly relies on the strength of the Balmer break, which is weaker in later type galaxies.

Similar conclusions are derived from the comparison between photometric redshifts of galaxies in the catalogue of AXAF and the redshifts for the same galaxies measured in the COMBO-17 survey \citep{Wolf03} as shown in Fig~\ref{sp}.

\section{The SN sample}

In this section we describe how we searched for SNe in our galaxy sample and obtained their redshift distribution.
We first illustrate the detection of variable sources and the selection of SN candidates (Sect.~\ref{search}) then their confirmation and classification (Sect.~\ref{classification}).

\subsection{SN candidate detection and selection}\label{search}

In order to achieve a rapid, automatic detection of SN candidates we developed the $STRESS$ package, a collection of {\em IRAF} tasks and freely distributed software. After the standard data reduction (see Sect.~\ref{Gdetect}), the three jittered exposures of a given field are mapped to a common geometrical grid, properly scaled in intensity and stacked together.  From this {\em search}, a suitable archive image ({\em template}) obtained at a different epoch is subtracted, and the {\em difference} image is searched for variable sources. 

After the images have
been geometrically registered and photometrically scaled, an accurate match of the PSF of the two images is obtained using the {\em mrj\_phot} task from the {\em ISIS} package \citep{Alard} which derives a convolution kernel comparing selected sources from the two images.  The best results in the image subtraction are obtained when the two images have similar seeing. For this reason we maintained an on$-$line archive of {\em template} images with different seeing. 
We note however that all the selected SN candidates were still detected even when using a different {\em template} image. 

Variable sources are detected as positive residuals in the difference image using {\em SExtractor} with a source detection threshold of $3 \sigma$ above the background noise and a minimum of 10 connected pixels above the threshold. 
The choice of the detection threshold is a trade-off between detecting sources as faint as possible and limiting the number of spurious detections.
Indeed, the list produced by {\em SExtractor} is heavily contaminated by spurious sources due to imperfect removal of bright stars, cosmic rays, hot or dead pixels. In the catalogue obtained from the difference image, false detections outnumber real variable sources by a factor of $\sim 100$.  In addition, besides the  SN candidates the detection list contains other variable sources such as fast moving objects, variable stars, variable AGNs.

To clean up the detection list we used a custom$-$made ranking program which assigns a score to each source based on several parameters measured by {\em SExtractor} in the {\em search}, {\em template} and {\em difference} images: the $class\_star$, the full width at half maximum (FWHM), the distance of the residual with respect to the center of the associate galaxy (if any), and apparent magnitudes of the residual measured with different prescriptions. The scoring algorithm, tuned through extensive artificial star experiments, produces a sorted detection list. The final selection of trusted variable sources, about ten per image, was performed through visual inspection by an experienced observer.

In particular, the residuals of fast moving objects, which were not completely masked by our jittering strategy, have irregular shape in the stacked image and are easily recognized by comparison of the individual dithered images. 
Conservatively, we classified as variable stars those sources showing a stellar profiles both in the {\em search} and in the {\em template} images.  
However, it is impossible to distinguish AGNs from SNe that exploded near galaxy nuclei by examining images taken at one epoch only.  We labeled all variable sources closer than 0.5 arcsec to the host galaxy nucleus as SNAGN candidates and maintained them in the follow-up target list to reduce biases in the SN candidate selection as much as possible.
The role of AGN contamination for those SNAGN candidates without spectroscopic observations will be discussed in Sect.~\ref{classification}. Finally we classified as SN candidates those variables with a stellar profile in the {\em difference} image, no matter if they appeared projected on a (host) galaxy in the {\em search} and in the {\em template} images, or not. However, in the latter case
we excluded the candidate from our analysis, since we concentrate on SNe 
occurring in the galaxies within our specific sample.

To deal effectively with the large number of epochs and candidates, we developed a {\em MySQL}\footnote{MySQL is an open source database released under the GNU General Public License (GPL), {\tt http://www.mysql.com/}.} database with a web
interface \citep{Riello} to easily access the information remotely (e.g.
during observing runs). In particular the database stores:
the monitored fields, pointing coordinates, finding charts, log of observations (with seeing, exposure time, photometric zero point), all discovered variable sources, their identification chart and relevant parameters (sky coordinates, epoch, magnitude, stellarity index, etc.). 

At the end of our SN search programme we reviewed all variable sources recorded in our database to obtain a final classification based on all information gathered during our observing programme. A search engine allows us to identify multiple detections of the same source in different epochs and filters. This was used  for the identification of AGNs which in general show long term, irregular variability. Indeed, the relatively long duration of our search makes this approach fairly efficient for AGN removal.

\subsection{SN candidate confirmation and classification}\label{classification}

\begin{table*}
\centering
\caption{Spectroscopically confirmed SNe}\label{snspec}
\renewcommand{\footnoterule}{}
\begin{tabular}{lccccccc}
\hline\hline
Design. & R.A.(J2000.0) & Dec.(J2000.0) & type & z & field & band   & reference\\
\hline
1999ey$^\#$ & 00:58:03.43 & $-$27:40:31.1 & IIn & 0.093 & J1888   & V   & IAUC 7310\\
1999gt & 03:32:11.57 & $-$28:06:16.2 & Ia  & 0.274 & AXAF    & V   & IAUC 7346\\
1999gu & 03:33:00.20 & $-$27:51:42.7 & II  & 0.147 & AXAF    & V   & IAUC 7346\\
2000fc & 00:58:33.55 & $-$27:46:40.1 & Ia  & 0.420 & J1888   & V   & IAUC 7537\\
2000fp & 05:38:01.27 & $-$23:46:34.1 & II  & 0.300 & EisC    & V  & IAUC 7549\\
2001bd & 19:13:10.94 & $-$64:17:07.8 & II  & 0.096 & Field2  & V   & IAUC 7615\\
2001gf & 22:04:21.27 & $-$18:21:56.8 & Ia  & 0.200 & 22Z3    & V   & IAUC 7762\\ 
2001gg & 00:45:20.83 & $-$29:45:12.2 & II  & 0.610 & EisB    & V   & IAUC 7762\\
2001gh & 00:57:03.63 & $-$27:42:32.9 & II  & 0.160 & J1888   & V  & IAUC 7762\\
2001gi$^\#$ & 04:28:07.06 & $-$36:21:45.2 & Ia  & 0.200 & EF0427   &V & IAUC 7762\\
2001gj & 05:38:07.90 & $-$23:42:34.4 & II  & 0.270 & EisC    & V  & IAUC 7762\\
2001io & 00:45:53.64 & $-$29:26:11.7 & Ia  & 0.190 & EisB    & V  & IAUC 7780\\
2001ip & 03:31:13.03 & $-$27:50:55.5 & Ia  & 0.540 & AXAF    & V  & IAUC 7780\\
2002cl & 13:44:09.94 & $-$00:12:57.8 & Ic  & 0.072 & 13Z3    & V   & IAUC 7885\\
2002cm & 13:52:03.67 & $-$11:43:08.5 & II  & 0.087 & WH23 & V   & IAUC 7885\\
2002cn & 13:53:21.24 & $-$12:15:29.5 & Ia  & 0.302 &WHhz2 & V   & IAUC 7885\\
2002co$^\#$ & 14:10.53.04 & $-$11:45:25.0 & II  & 0.318 & WH27 &V  & IAUC 7885\\
2002du & 13:53:18.28 & $-$11:37:28.8 & II  & 0.210 & WH23 & V  & IAUC 7929\\
2004ae & 04:28:17.89 & $-$36:18:55.0 & II  & 0.480 &  EF0427 & R   & IAUC 8296\\
2004af & 05:38:03.91 & $-$23:59:00.2 & Ic  & 0.056 & EisC    & R &   IAUC 8296\\
2004ag & 09:51:01.49 & $-$20:50:37.5 & II  & 0.362 &  EisD   & R &  IAUC 8296\\
2004ah & 10:45:47.41 & $-$00:06:58.1 & Ia  & 0.480 & 10Z2    & R &  IAUC 8296\\
2004ai & 13:54:26.09 & $-$12:41:15.9 & Ic  & 0.590 & WHhz2& R &   IAUC 8296\\
2004aj$^\#$ & 14:20:37.58 & $-$12:24:14.4 & Ia  & 0.247 & WH31 & R &   IAUC 8296\\
2004cd$^\#$ & 13:44:50.70 & $-$00:03:48.3 & Ia  & 0.500 & 13Z3    & R &  IAUC 8352\\
2004ce & 13:54:45.93  & $-$12:14:36.3 & Ia  & 0.465 & WHhz2& R &  IAUC 8352\\
2004cf & 14:11:05.77 & $-$11:44:09.4 & Ib/c& 0.247 & WH27 & R &   IAUC 8352\\
2004cg & 14:11:23.71 & $-$12:01:08.4 & II  & 0.264 &  WH27& R &   IAUC 8352\\
2004dl & 14:11:11.50 & $-$12:02:36.9 & Ia  & 0.250 & WH27 & R & IAUC 8377\\
2004dm & 22:43:21.36 & $-$40:19:46.7 & Ib  & 0.225 & EisA1   & R & IAUC 8377\\
2005cq$^\#$ & 09:52:00.48 & $-$20:43:26.5 & Ia  & 0.310  & EisD    & R  & IAUC 8551\\
\hline
\end{tabular}

\# - Not included in the rate computation because the host galaxy has $R>21.8$.
\end{table*}

Spectroscopic observations were planned for all SN and SNAGN candidates but could not be completed due to limited telescope time and, in some cases, poor weather conditions. In general a higher priority was given to candidates flagged as SN. 
Spectrum extraction and calibration were performed using standard {\em IRAF} packages.
Special care was devoted to removal of background contamination through a detailed analysis of the intensity profile along the slit. For the nuclear candidates it is impossible to separate the variable source from the background using this information only. In such cases, first the sky emissions was subtracted, then the spectrum was extracted by integrating all the light along the slit. A best fit galaxy spectral template extracted from the STSci Database of UV-Optical Spectra of Nearby Quiescent and Active Galaxies \footnote{{\tt http://www.stsci.edu/ftp/catalogs/nearby\_gal/sed.html}} was then subtracted from the candidate spectrum.

The classification of SN candidate spectra was facilitated by the use of {\em Passpartou}, a software package that automatically classifies the candidate spectrum after comparison with all the SN spectra in the archive of the Padova SN group \citep{Avik}.  

We could obtain direct spectroscopic observations only for 38 candidates ($\sim 40\%$ of the total): 31 were confirmed to be SNe whereas 7 turned out to be AGN.  
In this paper we use only the information on SN type and redshift of the host galaxy while in a future paper we will give more details on the spectroscopic data analysis.

The properties of the 31 confirmed SNe are summarized in Table \ref{snspec}: 25 have been observed in our galaxy sample (9 SNe~Ia and 16 SNe~CC) whereas 6 SNe discovered in galaxies with $R>21.8$ mag were excluded from the following analysis. The SN redshifts range from $z=0.056$ to $z=0.61$.

As mentioned in Sect.\ref{obs}, we carried out a complementary observing programme to estimate redshift and spectral type of SN host galaxies and to better constrain contamination of AGNs in the SNAGN candidate sample. 
 We remark that the higher priority given to SN candidates for immediate follow-up introduces a bias against SNAGN candidates. Such bias is minimized by this late follow-up which mostly targets the host galaxies of the SNAGN candidates.
We found that, out of 42, 22 host galaxies showed an active nucleus. 
All the candidates occurring in these galaxies were removed from the candidate list.  Actually we have to notice that with our approach a SN occuring in a galaxy with an active nucleus will be discharged if only late spectroscopy is available.
In Table \ref{snhost} we report coordinates, redshifts, search filter and classifications of the remaining 20 SN candidates with host galaxy spectroscopy.

\begin{table}
\caption{SN candidates with host galaxy spectra}\label{snhost}
\begin{tabular}{lccccc}
\hline\hline
Design. & R.A.(J2000.0) & Dec.(J2000.0) & z & band \\
\hline
03Z3$\_$B &03:39:27.76 &$-$00:01:08.8 &0.240 & V\\ 
03Z3$\_$F &03:40:36.80 &$-$00:06:17.8 &0.247 & V\\ 
03Z3$\_$H &03:40:09.43 &$-$00:10:20.6 &0.282 & V\\ 
03Z3$\_$J &03:40:37.11 &$-$00:05:22.5 &0.180 & V\\ 
10Z2$\_$B &10:45:42.76 & 00:00:27.9 &0.295 & V\\ 
10Z2$\_$D &10:47:40.94 &$-$00:13:52.6 &0.350 & V\\ 
EF0427$\_$G &04:30:21.93 &$-$36:06:49.9 &0.359 & V\\ 
EF0427$\_$I &04:30:10.73 &$-$36:25:47.4 &0.420 & V\\ 
EF0427$\_$J &04:29:20.47 &$-$36:32:16.7 &0.185 & V\\
EF0427$\_$O &04:28:10.56 &$-$36:13:54.4 &0.450 & V\\ 
EF0427$\_$Q &04:28:17.09 &$-$36:09:54.9 &0.394 & V\\ 
EisA1$\_$B &22:44:01.94 &$-$40:03:21.7 &0.219 & V\\ 
EisA1$\_$C &22:44:25.26 &$-$40:20:01.9 &0.215 & V\\ 
EisA1$\_$J &22:44:02.82 &$-$40:02:05.2 &0.180 & V\\ 
EisA1$\_$Q &22:42:37.87 &$-$39:57:54.4 &0.380 &V\\  
EisA1$\_$R &22:42:13.27 &$-$40:03:19.9 &0.530 &V\\ 
EisC$\_$A &05:38:45.67 &$-$23:44:41.0 &0.643 &V\\ 
EisC$\_$C &05:38:06.55 &$-$23:36:57.1 &0.133 &V\\  
EisC$\_$G &05:37:52.11 &$-$23:38:14.4 &0.156 &V\\ 
WH27$\_$B  & 14:11:02.77 & $-$11:49:06.2 &0.485 &V \\
WHhz2$\_$A & 13:53:58.73 & $-$12:43:04.8 &0.444 &V \\
WHhz2$\_$B &13:54:25.27 &$-$12:35:42.9 &0.322&V\\
\hline
\end{tabular}
\end{table}

Even after this complementary analysis, 13 SN and 28 SNAGN candidates still
remained  without direct or host galaxy spectroscopy.  The list of these remaining candidates is reported in Tab.~\ref{sncand}.

\begin{table}
\caption{Candidates without direct or host galaxy spectroscopy}\label{sncand}
\begin{tabular}{lcccccc}
\hline\hline
Design. & R.A.(J2000.0) & Dec.(J2000.0) & z & band & class.\\
\hline
03Z3$\_$Q    & 03:38:26.39 & $-$00:04:53.1 & 0.30 &V & snagn\\
10Z2$\_$A    & 00:47:06.51  & 00:00:39.6     &0.52 &R & snagn\\ 
13Z3$\_$R    & 13:44:01.94 & 00:00:03.9  & 0.35 &R & snagn\\    
13Z3$\_$X    & 13:45:22.55 & 00:03:11.2  &0.18 &R & snagn\\     
13Z3$\_$AB   & 13:44:31.26 & 00:00:57.5  & 41&R & sn\\       
13Z3$\_$AC   & 13:45:11.26 & $-$00:12:27.4 &0.43 &R & snagn\\          
22Z3$\_$I       & 22:06:03.26 & $-$18:28:10.8 &0.42 &R & sn\\      
22Z3$\_$J      & 22:04:42.12 & $-$18:36:38.3 & 0.19&R & sn\\   
22Z3$\_$K     & 22:04:35.87 & $-$18:45:45.0 &0.42 &R & snagn\\    
22Z3$\_$N     & 22:04:49.66 & $-$18:45:27.3 & 0.14&R & snagn\\      
22Z3$\_$R     & 22:04:03.87 & $-$18:35:55.2 & 0.22&R & sn\\  	          
AXAF$\_$B     & 03:32:31.16 & $-$28:04:43.8 &0.13 &V & sn\\ 
AXAF$\_$D     & 03:31:28.98 & $-$28:10:26.1 &0.42 &V & snagn\\ 	           
AXAF$\_$F     & 03:33:05.29 & $-$27:54:09.2 &0.17 &V & snagn\\           
AXAF$\_$G     & 03:32:39.23 & $-$27:42:57.5 &0.47 &R & sn\\          
AXAF$\_$N     & 03:32:47.68 & $-$28:07:41.1 &0.18 &R & sn\\ 
EF0427$\_$B  & 04:29:07.28 & $-$36:28:01.8 & 0.50&V & sn\\ 
EF0427$\_$F  & 04:29:49.31  & $-$36:33:55.2 & 0.24&V & snagn\\          
EF0427$\_$M & 04:27:57.15 & $-$36:30:52.9 &0.74 &V & snagn\\                     
EisA1$\_$E   & 22:43:44.86 & $-$40:05:50.6 &0.17 &V & snagn\\ 
EisA1$\_$I   & 22:44:48.37 & $-$40:03:42.3 & 0.17&V & snagn\\   
EisA1$\_$O   & 22:42:01.73 & $-$40:21:17.8 &0.25 &V & snagn\\                 
EisA1$\_$AC  & 22:44:21.82 & $-$40:08:21.4 &0.52 &R & sn\\               
EisB$\_$B    & 00:45:09.35 & $-$29:28:33.2 &0.13 &V & snagn\\          
EisB$\_$P    & 00:44:14.57 & $-$29:39:08.1 & 0.17&V & snagn\\
EisB$\_$S    & 00:45:51.29 & $-$29:52:10.4 & 0.43&V & snagn\\ 
EisC$\_$H    & 05:39:28.83 & $-$23:38:05.7 &0.36 &V & sn\\   
EisC$\_$J    & 05:38:01.65 & $-$23:51:40.7 &0.20 &V & snagn\\ 
EisC$\_$K    & 05:38:28.78 & $-$23:46:29.4 &0.80 & V & snagn\\   
EisC$\_$O    & 05:39:28.51 & $-$24:00:05.4 &0.80 & V & snagn\\ 
EisC$\_$Q    & 05:37:50.49 & $-$23:36:55.2 &0.14 & R & sn\\ 
EisD$\_$F    & 09:51:12.26 & $-$20:56:14.0 &0.16 &R & snagn\\          
Field2$\_$C  & 19:11:16.62 & $-$64:17:32.3 &0.16 &V & snagn\\ 
Field2$\_$E  & 19:10:35.43 & $-$64:21:36.0 &0.28 &R & sn\\         
J1888$\_$J   & 00:58:41.35 & $-$27:50:38.1 & 0.30 &V & sn\\   
J1888$\_$M   & 00:57:05.34 & $-$27:45:57.7 &0.63 &V & snagn\\            
WH23$\_$A    & 13:53:18.79 & $-$11:40:29.7 &0.16 &V & snagn\\
WH23$\_$E    & 13:52:41.23 & $-$11:51:24.7 &0.21 &R & snagn\\ 
WH27$\_$N    & 14:11:47.81 & $-$11:35:25.6 & 0.44&R & snagn\\                      
WHhz2$\_$Q   &13:54:31.46 & $-$12:18:22.5 &0.67 &R & snagn\\      
WHhz2$\_$T   & 13:55:08.90 & $-$12:18:38.3 & 0.23&R & snagn\\        
\hline
\end{tabular}
\end{table}

For all these candidates the redshift of the host galaxy was estimated using the {\em photo-z} technique.
To avoid a possible contamination by the SN candidate in measuring the host galaxy colours we produced deep images stacking only those frames obtained
6 -12 months earlier than the epoch of discovery of the candidate.

\begin{figure}
\resizebox{\hsize}{!}{\includegraphics{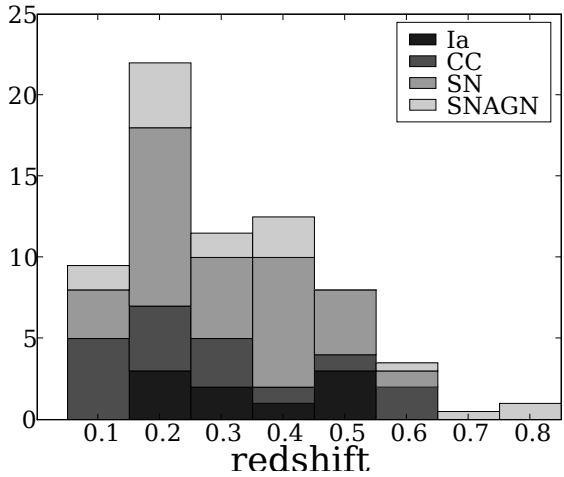}}
   \caption{The redshift distribution of SN candidates with (Ia and CC) and without spectroscopic confirmation (SN and SNAGN).}
    \label{sndist}
\end{figure}

In summary the analysis presented here is based on 86 objects: 25 spectroscopically confirmed SNe, 33 SN candidates and 28 SNAGN candidates.
The redshift distribution of candidates peaks at $z\sim 0.2-0.3$ and extends up to $z\sim0.6$ as shown in Fig ~\ref{sndist}. 
Combining information from the long term variability history, direct and host galaxy spectroscopy we found that $\sim 50\%$ of the variable sources originally classified SNAGN were actually AGN.
Hereafter, in the analysis of SN statistics, we assigned a weight of 0.5 to all SNAGN candidates with no spectroscopic observations.
The unclassified SN and SNAGN candidates were distributed among type Ia and CC based on the observed fractions in each redshift bin for the SN subsample with spectroscopic confirmation.
It turns out that, with respect to the total number, SNe~Ia are $27\pm18$\%, $47\pm21$\% and $63\pm36$\% at redshift $z=0.1,0.3,0.5$ respectively.
We quantify the effect of misclassification for the candidates without spectroscopic observations by performing Monte Carlo simulations as illustrated in Sect.~\ref{errors}.

\section{The control time of the search}\label{ctime}

The time interval during which a SN occurring in a given galaxy can be detected (the control time) depends mainly on the SN detection efficiency (Sect.~\ref{efficiency}), the SN light curve (Sect.~\ref{lightcurves}) and the amount of dust extinction (Sect.~\ref{extinction}). 
In this section we describe the different ingredients needed to calculate the control time of each single observation and the recipe to obtain the total control time of our galaxy sample (Sect.~\ref{controltime}).

\subsection{The search detection efficiency}\label{efficiency}

The SN detection efficiency depends mainly on the survey strategy, instrumental set up and observing conditions. For a given instrumental set up and exposure time, the detection efficiency curve as a function of the SN candidate apparent magnitude was estimated via Monte Carlo simulations.

In each simulation artificial point-like sources of different magnitudes were added to an image which was then searched for variable sources using the same software as in the actual search. The detection efficiency at a given magnitude is computed as the ratio between the number of recovered and injected artificial sources.

\begin{figure}
\resizebox{\hsize}{!}{\includegraphics{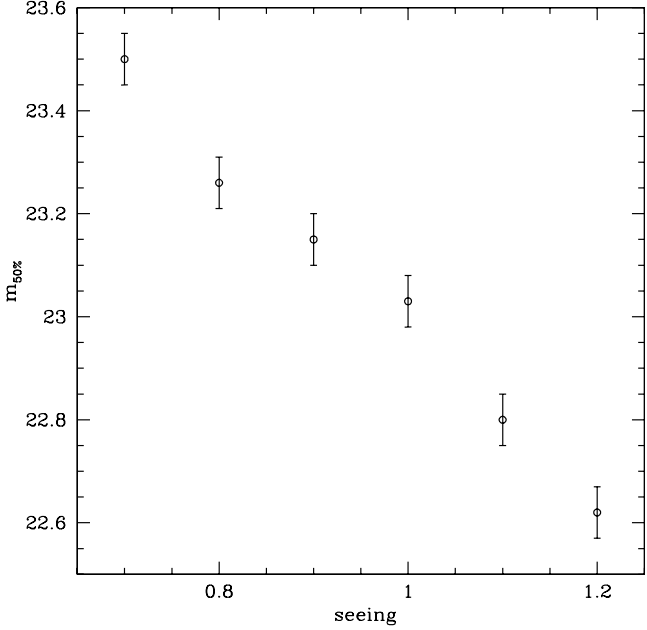}}
\caption{The magnitude at which the detection efficiency is $50\%$, $m_{50\%}$, in $R$ band as function of seeing.}
\label{deteff}
\end{figure}

Artificial point-like sources were generated using the {\em IRAF} {\em Daophot} package with the PSF determined from a number of isolated stars on each chip of each image. The artificial stars were then placed on the galaxies (one source per galaxy) in a random position with respect to the center, but following the the luminosity profile. 
Detailed simulations for a few selected fields allowed us to probe the most relevant parameters affecting the detection efficiency: the search image characteristics (observing conditions and effective search area), the search process (choice of template image and selection criteria) and the SN position in the host galaxy.

In particular we found that, for a given seeing, differences in the sky transparency do not affect the shape of the efficiency curve. 
The magnitude at which the efficiency is $50\%$ ($m_{50\%}$) becomes  fainter when the sky transparency improves, and a variation of photometric zero point translates directly into a change in $m_{50\%}$.
 
After accounting for the differences in sky transparency, a strong dependence of the detection efficiency on seeing is still present (Fig.~\ref{deteff}). We found that $m_{50\%}$ is $\sim1$ mag brighter when the seeing degrades from 0.7 arcsec to 1.2 arcsec, as shown in Fig.~\ref{deteff}. If the seeing is worse than 1.3-1.4 arcsec then $m_{50\%}$  decreases more rapidly.

The detection efficiency for different fields is very similar except for minor variations due to different effective search areas caused by the presence different numbers of bright stars (which were masked).
This is also the reason why the detection efficiency is always lower than $100\%$, even for bright sources.

In  order to reduce the overhead of determining the detection efficiency for each individual observations of a given field we performed an extensive set of Monte Carlo simulations for only one field, covering a wide range of source magnitudes and seeing, thereby generating a set of template efficiency curves. Then, for each individual observation of the others fields, we performed only few simulations to determine the shift to apply to the appropriate template curve.

Possible spurious effects caused by the choice of the {\em template} image were tested by performing the same analysis based on three other images with different seeing. We found that, in the range $21-23$ mag, $95\%$ of the sources are recovered in all cases, while, at fainter magnitudes, this fraction quickly decreases. 

As we mentioned in Sect.~\ref{search}, the final selection of SN candidates was based on visual inspection. We verified that the detection efficiency is independent of the observer that carried out the visual inspection by comparing the list of candidates produced by different observers in several simulations. 
Finally,  we found that the detection efficiency depends very little on the position of the source with respect to the galaxy and with galaxy brightness except for nuclear candidates in very bright galaxies.

This analysis of the detection efficiency was carried out both on $R$- and $V$-band images with very similar results after considering the zero-point differences.
 
\subsection{SN light curves}\label{lightcurves}

The observed light curve in the band $F$ of a given SN type at redshift $z$ is given by:

\begin{eqnarray}\label{lc}
  m^{\rm SN}_{F}(t,z) &= & M^{\rm SN}_{B}(0)+ \Delta M^{\rm SN}_B(t')  + K^{\rm SN}_{BF}(z,t')
  \nonumber \\ 
                                    & & +  A_F^G+ A_B^h + \mu(z) 
\end{eqnarray}
where:
\begin{itemize}
\item  $M^{\rm SN}_{B}(0)$ is the SN absolute magnitude at maximum in the $B$ band, 
\item $\Delta M^{\rm SN}_B(t')$ describes the light curve relative to the maximum,
\item $K^{\rm SN}_{BF}(z,t')$ is the K-correction from the $B$ to the $F$ band, ($V$ or $R$),
\item $ A_F^G$ and $ A_B^h$ are the Galactic and host galaxy absorption respectively,
\item $\mu(z)$ is the distance modulus for the redshift $z$ in the adopted cosmology.
\end{itemize}

The light curve is translated into the observer frame by applying the time dilution effect to the rest frame SN phase $t' =t/(1+z)$.
Four basic SN types were considered: Ia, IIL, IIP, Ib/c.
We adopted the same templates for the SN $B$-band light curves as in \citet{Cappellaro97} and account for the observed dispersion of SN absolute magnitudes at maximum light by assuming a normal distribution with the mean $M_{B,0}$ and $\sigma$ listed in Tab.~\ref{tababsmag}. 
In this table we also report the average absorption $<A_B>$ measured in a sample of nearby SNe.

\begin{table}
\centering
\caption{Adopted $B$ band, absolute magnitude at maximum for different SN types.}\label{tababsmag}
\begin{tabular}{lccccl}
\hline\hline
  SN type  &  $M_{B,0}$  & $\sigma_{B,0}$  & $<A_B>$ & Ref. \\
\hline
      Ia        &  $-19.37$    &       0.47             & 0.4~~       &\citet{Wang} \\
      Ib/c     &  $-18.07$    &       0.95             &  0.7~~       &\citet{Richardson06}\\
      IIP       &  $-16.98$    &       1.00             &  0.7$^\#$      &\citet{Richardson02}\\ 
      IIL       &  $-18.17$    &       0.53             &  0.7$^\#$      &  ~~~~~~"~~~~~~~~~~" \\
\hline
\end{tabular}

\# - \citet{Nugent}
\end{table}

Finally, Galactic extinction is taken into account using the \citet{Schlegel} extinction maps, whereas the correction of the host galaxy extinction is estimated through a statistical approach, described in Sect.~\ref{extinction}.

\subsection{$K$ correction}\label{kcor}

For a given filter, $K$-correction is the difference between the magnitude measured in the observer and in the source rest frame which depends on the redshift and SED of the source. In principle, if the source redshift is known it is possible to design a filter properly tuned to minimise the need for a $K$-correction. In most practical cases,  however, sources are distributed in a wide range of redshifts and this approach is inapplicable.

A detailed discussion of the formalism and application of $K$-correction to the case of type Ia SNe can be found in \citet{Nugent02}. SN~Ia have received a special care as their use as cosmological tools requires high precision photometry.
Actually, for computing the control time, we can tolerate much larger uncertainties. However, we need to obtain the K correction also for type Ib/c and type II SNe, which have not been published so far.

In general, there are three main problems  for an accurate estimate of the $K$-correction of SNe: $i)$ the SED rapidly evolves with time $ii)$ SNe, in particular type Ib/c and type II, show large SED diversities, $iii)$ the UV spectral coverage, which is needed to compute optical $K$-correction at high redshift, is available only for few SNe, and typically only near maximum light.

We addressed these issues trying to use all available informations on SN SED, in particular:

\begin{enumerate}
\item we selected a sample of spectral sequences of different SN types with good S/N retrieved from the Asiago-Padova SN archive. The sample includes
the type Ia SNe 1990N, 1991T, 1991bg, 1992A, 1994D, 1999ee, 2002bo, the type II
SNe 1979C, 1987A, 1992H, 1995G, 1995AD, 1996W, 1999el, 2002GD, 2003G  and the type Ib/c SNe 1990I, 1997B, 1997X, 1998bw. 
\item $UV$ spectral coverage was secured retrieving SN spectra from the ULDA SN spectra archive (SN~Ia 1990N phase -9, 1981B phase 0) of the International Ultraviolet Explorer \citep{ULDA} and from the HST archive (SN~Ia 1992A at phase +5, +45, SN~Ib/c 1994I phase +7d, SN~1999em phase 5d) . These were combined with optical and IR spectra to obtain the SED in a wide range of wavelengths.
\item  we included models for standard SNIa at phase -7, 0, +15 ~d provided by \cite{mazzali} and models of the SNII 1999em at phase 0,+9,+11,+25 ~d, courtesy of \cite{baron}.
\end{enumerate}

The observed and model spectra were shifted at different redshifts and synthetic photometry in the required bands was obtained using the {\em IRAF} package {\em synphot}. 
Eventually, we derived the $K^{\rm SN}_{BV}$ and $K^{\rm SN}_{BR}$ corrections for each SN type as a function of light curve phase with a step of 0.05 in redshift.
Examples of the measured $K$ correction as a function of the SN phase are shown in on line Figs.~\ref{kbv01}-\ref{kbr06}.

\onlfig{1}{
\begin{figure}
\resizebox{\hsize}{!}{\includegraphics{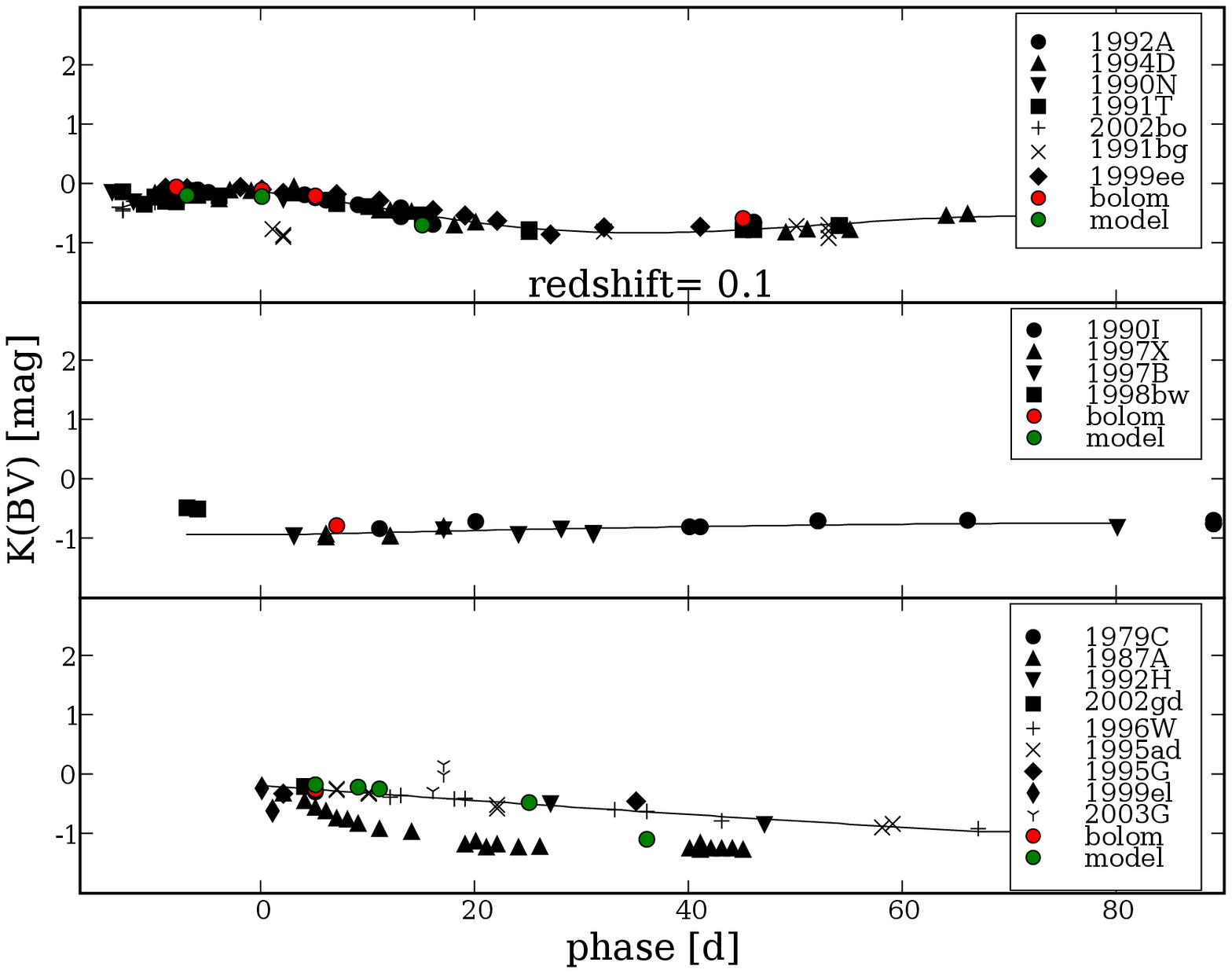}}
\caption{K correction as a function of SN phase from the observed $V$ band to the rest frame $B$ band for redshift 0.1. Top panel shows the K(BV) for a type Ia SN, mid panel for a type Ib/c SN and bottom panel for a type II SN. In all cases the best fit has been obtained neglecting peculiar SNe as SN 1991bg for type Ia SN and SN 1987A for type II SN. }
\label{kbv01}
\end{figure}
}
\onlfig{14}{
\begin{figure}
\resizebox{\hsize}{!}{\includegraphics{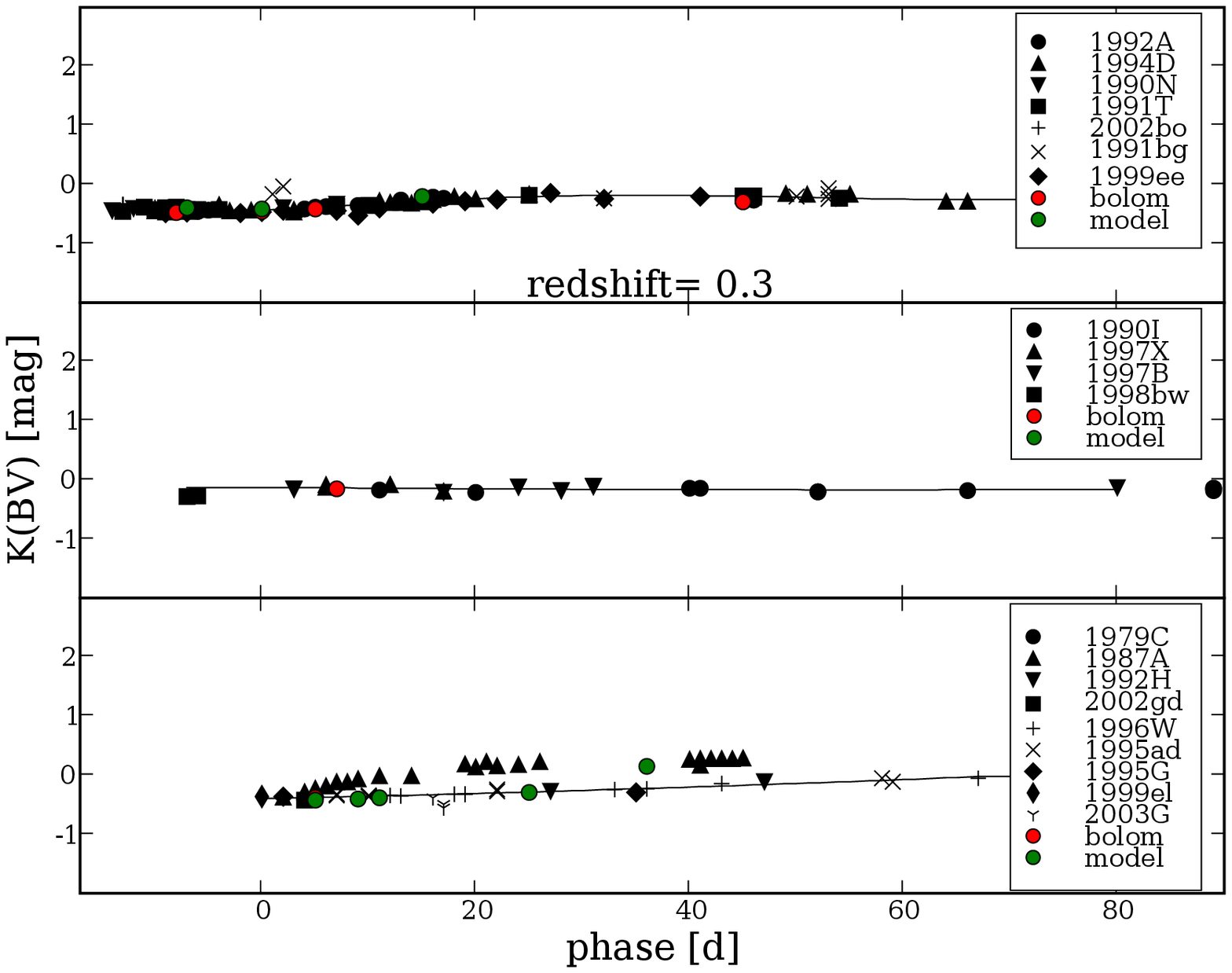}}
\caption{K(BV) correction for redshift 0.3.}
\label{kbv03}
\end{figure}
}
\onlfig{15}{
\begin{figure}
\resizebox{\hsize}{!}{\includegraphics{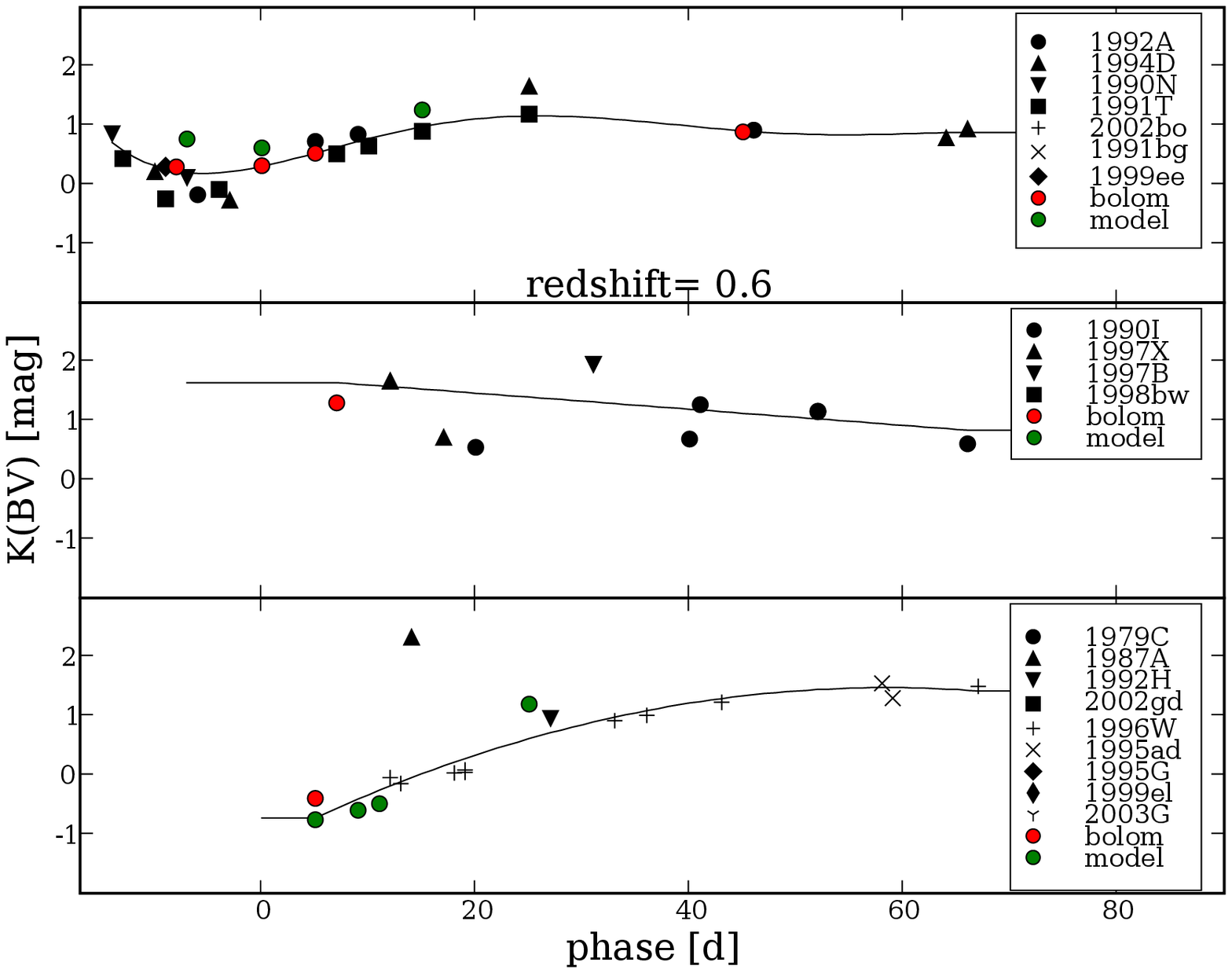}}
\caption{K(BV) correction for redshift 0.6.}
\label{kbv06}
\end{figure}
}
\onlfig{16}{
\begin{figure}
\resizebox{\hsize}{!}{\includegraphics{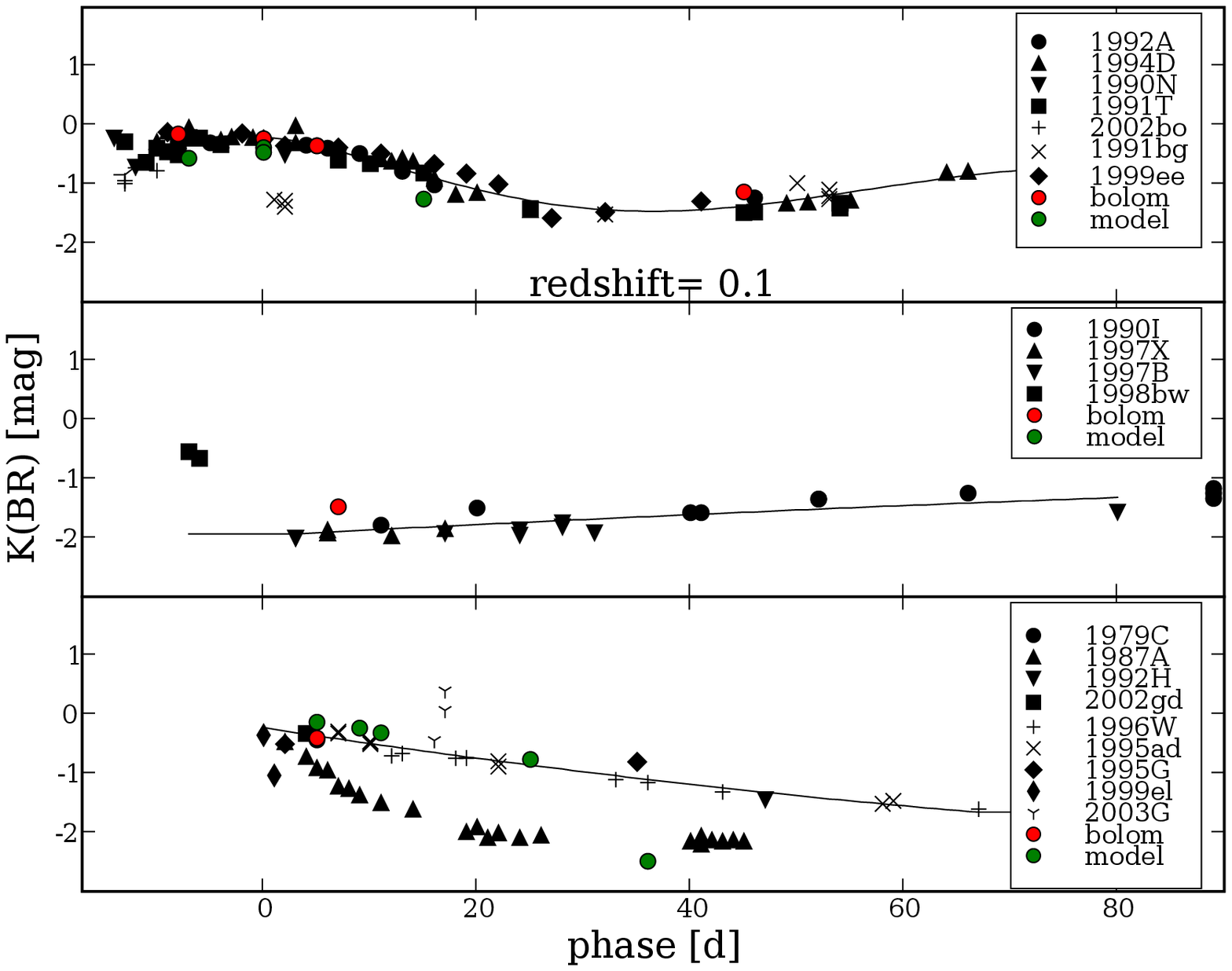}}
\caption{Same as Fig.1 for the K correction from the observed $R$ band to the rest frame $B$ band for redshift 0.1.}
\label{kbr01}
\end{figure}
}
\onlfig{17}{
\begin{figure}
\resizebox{\hsize}{!}{\includegraphics{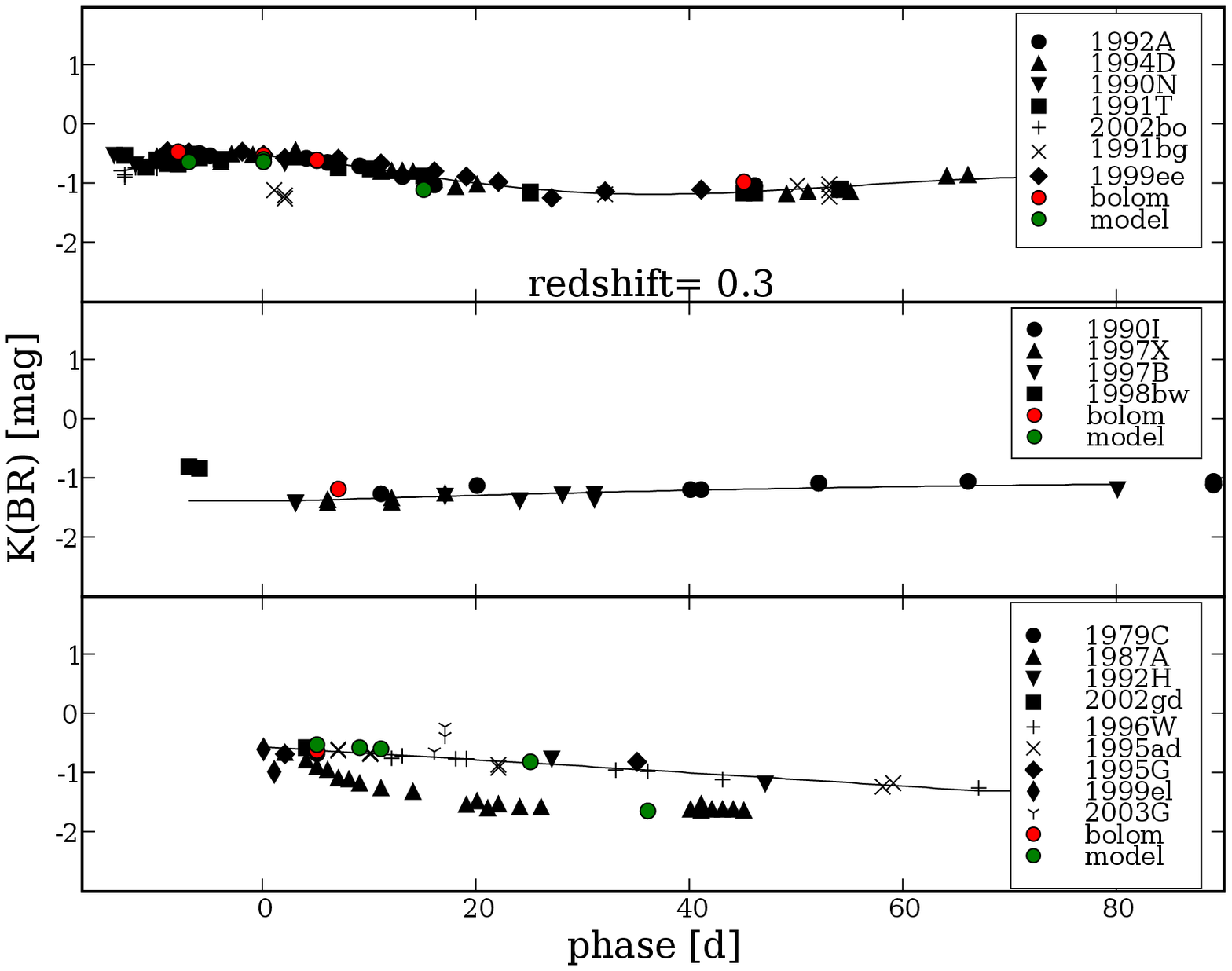}}
\caption{K(BR) correction for redshift 0.3.}
\label{kbr03}
\end{figure}
}
\onlfig{18}{
\begin{figure}
\resizebox{\hsize}{!}{\includegraphics{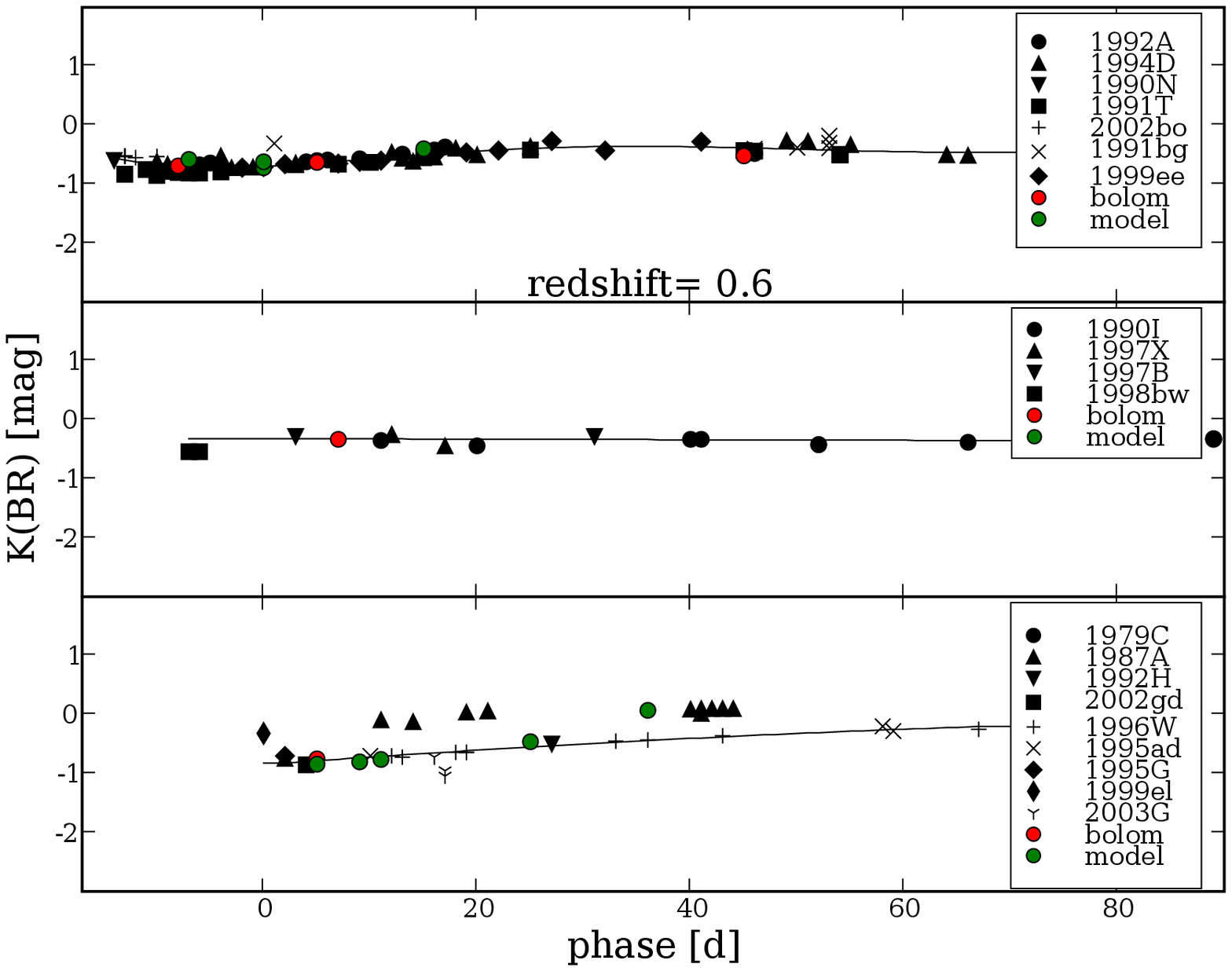}}
\caption{K(BR) correction for redshift 0.6.}
\label{kbr06}
\end{figure}
}
Uncertainties due to variances of the SN spectra, uncertain extinction correction and/or incomplete temporal/spectral coverage can be minimized by choosing a proper filter combination. Indeed as it can be seen in on-line Fig.~\ref{kbv03}, for the average redshift of our search, $\overline{z} \sim0.3$, $K^{\rm SN}_{BV}$ is almost independent on the SN type and phase.

\subsection{Host galaxy extinction}\label{extinction}

The host galaxy extinction correction is the most uncertain ingredient in the 
estimate of the SN rate.
In the local Universe \citet{Cappellaro99} relied on an empirical correction as a function of the host galaxy morphological type and inclination. This approach cannot be applied to galaxies of our sample because the relevant data are not available. Hence we resorted to a statistical approach, based on the modelling of SN and extinction distribution in galaxies. 

In short, following the method described in \citet{RP05}, we performed a number of Monte Carlo simulations where artificial SNe were generated with a predefined spatial distribution function and were seen from uniformly distributed lines of sight. Integrating the dust column density along the line of sight for each SN we derived the total optical depth and the appropriate extinction was applied to the SN template spectrum. Repeating a number of simulations we obtained the expected distribution of SN absorptions.

With respect to \citet{RP05}, we included in the modelling also CC SNe. Such extension was straightforward: we added spectral templates for CC SNe, and adopted a reasonable spatial distribution for the simulated CC SNe. Since CC SNe are thought to originate from massive progenitors, the spatial distribution is assumed to be concentrated in dust-rich regions along the spiral arms.

\begin{table}
\centering
\caption{Parameters of our modelling of the host extinction for the four SN types considered in the simulations. The last column of the table provides the reference from \citet{RP05} where the meaning of each parameter is explained in detail.}\label{tabmodels}
\begin{tabular}{lccccl}
\hline\hline
Parameter &  Ia  & Ib/c  & IIP & Dust & Ref.\\
\hline
$B/T$ & 0.5 & 0 & 0 & - & \S\ 2.1\\
$r^b$ & 1.8 & - & - & - & Eq. (1)\\
$n$ & 8 & - & - & - & \S\ 2.1.1\\
\hline
$r^d$ & 4.0 & 4.0 & 4.0 & 4.0 & Eq. (2)\\
$z^d$ & 0.35 & 0.35 & 0.35 & 0.14 & Eq. (2)\\
$n$ & 6 & 6 & 6 & 6 & \S\ 2.1.2\\
$m$ & 6 & 6 & 6 & 6 & \S\ 2.1.2\\
$N_a$ & 0 & 2 & 2 & 2 & Eq. (3)\\
$w$ & - & 0.2 & 0.2 & 0.4 & Eq. (3)\\
$p$ & - & 20 & 20 & 20 & Eq. (3)\\
\hline
\end{tabular}
\end{table}

The parameters for SN and dust spatial distributions used in the simulations are summarized in Tab. \ref{tabmodels}.  The top rows (1--3) of the Table show the value of the parameters that describe the bulge component, that is the bulge to total luminosity ratio ($B/T$), the scale and the truncation radius of SN distribution ($r^b$ and $n$ respectively). The bottom half of the table (rows 4--10) lists the parameters for the disk component of both SN and dust, namely the scale-length and scale-height of the disc ($r^d$,$z^d$), the respective truncation radii ($n$,$m$) and the spiral arm perturbation parameters. The reader should refer to \citet{RP05} for a more detailed description of $N_a$,$w$,$p$ parameters.

The two other parameters governing the dust properties are the total to selective extinction ratio $R_V$ and the total optical depth along the galaxy rotation axis $\tau(0)$, which provides a convenient parameterization for the total amount of dust in the galaxy. For the former we adopted the canonical value $R_V=3.1$.  For the latter we considered two scenarios: a {\em standard} extinction scenario with $\tau(0)=1.0$ for both SN types and a {\em high} extinction scenario with $\tau(0)=5.0$ only for CC SNe.

For each SN type we ran a set of simulations with $10^5$ artificial  SNe covering the redshift range $z=0.05-0.80$ with a step in redshift of 0.05.
In order to speed-up the computation, rather than considering a library of SN spectra at different phases as in \citet{RP05}, we used only a single spectrum close to maximum light for each SN type. This is a reasonable approximation since the evolution of the SN SED with time has a negligible effect on the absorption (at least for the SN phases that are relevant for our search). 

Once a simulation for a given parameter set was completed, we derived the observed distribution of SN absorptions, and used it to compute the control time (cf. Sect.~\ref{controltime}). 
For $\tau(0)=1.0$, the average absorption from the modeling is $<A_B^h>=0.5$ mag for SNe~Ia and  $<A_B^h>=0.7$ mag for CC SNe, after excluding SNe with absorption larger than $A_B^h=3$ mag (about 8$\%$ of SNe~Ia and 12$\%$ of CC SNe in this model, which are severely biased against in typical optical searches).  
We note that these average absorptions are consistent with the observed ones, reported in Tab.~\ref{tababsmag}.
For $\tau(0)=5.0$ the average model absorption is $<A_B^h>=0.8$ mag for CC SNe with $A_B^h<3$ mag (about 50$\%$ of the CC SNe)  which is still consistent with the observed value in the local Universe.

Our correction for dust extinctions is meant to provide a reasonable estimate
of this effect.
 A more realistic model should consider the different amount and distribution of the dust in each galaxy type.  This is far beyond the scope of the current analysis.

\subsection{Control time calculation}\label{controltime} 

We now detail how all the ingredients described in the previous sections have been combined to compute the control time of our galaxy sample. 
First, for a given SN type and filter $F$, the control time of the $i$-observation of the $j$-galaxy was computed as:

\begin{equation} 
CT^{\rm SN,F}_{j,i} = \int{\tau^{\rm SN,F}_{j}(m) \,\epsilon^{\rm F}_{i} (m)\,{\rm d}m}
\end{equation}
where $\tau^{\rm SN,F}_{j}(m)$ is the time spent by the SN in the magnitude range $m$ and $m+{\rm d}m$ and $\epsilon^{\rm F}_{i} (m)$ is the detection efficiency. 
More specifically, we convolved the distribution of the absolute magnitude at maximum ($M^{\rm SN}_{B}(0)$) and of the absorption due to host galaxy extinction ($A_B^h$), so as to determine the distribution of the quantity $M^{\rm SN}_{B}(0)+A_B^h$ appearing in Eq.~\ref{lc}, and computed $\tau^{\rm SN,F}_{j}(m)$ as a weighted average of the individual times over this combined distribution.

Then, the total control time $CT^{\rm SN,F}_j$ of the $j$-galaxy was computed by summing the contribution of individual observations.
If the temporal interval elapsed since the previous observation is longer than the control time, that contribution is equal to the control time of the observation, otherwise it is equal to the interval of time between the two observations \citep{Cappellaro99}. 

The total control time of the galaxy sample
is obtained as the $B$ band 
luminosity weighted ($\overline {CT}^{\rm SN,F}_j$) average of the individual galaxies. 

Since we merged all CC subtypes, including type Ib/c IIP and IIL, the control time for CC SNe is computed as follows:

\begin{equation}\label{eqctcc}
\overline{CT}^{\rm CC}_j = f_{\rm Ib/c}\, \overline{CT}^{\rm Ib/c}_j +  f_{\rm IIL}\, \overline{CT}^{\rm IIL}_j +  f_{\rm IIP}\, \overline{CT}^{\rm IIP}_j
\end{equation}

where the relative fractions of the different CC subtypes is assumed to be constant with redshift and equal to that observed in the local Universe, namely 20$\%$ of Ib/c and 80$\%$ of II \citep{Cappellaro99}, out of which 35$\%$ are IIL and 65$\%$ are IIP events \citep{Richardson02}.

In order to illustrate the role of the host galaxy extinction, we
calculated the CC SN rate by adopting the control time both for a {\em standard} ($\tau(0)=1.0$), and for a {\em high} ($\tau(0)=5.0$) extinction scenario.  

\section{SN rates}\label{snrate}

In this section we describe our approach to estimate SN rate, present our results (Sect.~\ref{snratecalc}) and discuss their uncertainties (Sect.~ \ref{errors}). We also compare our measurements to those obtained in the local Universe (Sect.~\ref{snrateev}).

\subsection{SN rate estimate}\label{snratecalc}
The SN rate at redshift $z$ is given by the ratio between the number of observed SNe and the control time of the monitored galaxies at that redshift.
Since our SN sample spans a wide redshift range ($0.056-0.61$), we can obtain an observational constraint on evolution of the rate by analysing the redshift
distribution of the events.

In analogy with \citet{Pain02}, and following analysis \citep{Dahlen04,Galyam04}, we adopt the following power law for the reshift dependence of the rate:

\begin{equation}
\label{rate1}
r^{\rm SN,F}(z) =r^{\rm SN,F}(\overline{z}) \left( \frac{ 1+z }{ 1+ \overline{z}} \right) ^{\alpha^{\rm SN,F}} 
\end{equation}

where $r(\overline{z})$ is the rate at the mean redshift of the search, $\overline{z}$, and $\alpha$ is the evolution index. Writing the rate evolution in terms of its value at the average redshift of the search reduces the correlation between the two free parameters in Eq.~(\ref{rate1}).

The reference redshift $\overline{z}$ of the search is computed as the weighted average of the galaxy redshifts  with weights given by the respective control times:

\begin{equation} 
\overline{z}^{\rm SN,F} = \frac{\sum_{j=1}^n z_j \,
\overline{CT}_j^{\rm SN,F}}{\sum_{j=1}^n \overline{CT}_j^{\rm SN,F}}
\end{equation}

We obtained:
\begin{itemize}
\item $\overline{z}^{\rm Ia}=0.30^{+0.14}_{-0.14}$
\item $\overline{z}^{\rm CC}=0.21^{+0.08}_{-0.09}$
\end{itemize}

Clearly, the lower $\overline{z}$ for CC SNe is due to their being, on average, intrinsically fainter than SNe~Ia; hence their control time at a given redshift is shorter. 

For a given SN type and filter F, the number of expected SNe is given by the expression:

\begin{equation}
N^{\rm SN,F}(z) = \frac{ r^{\rm SN,F}(z)}{(1+z)} \sum_{j=1}^n \overline{CT}^{\rm SN,F}_j(z) 
\end{equation}

where the sum is extended over the $n$ galaxies at redshift $z$ and the factor $1+z$ corrects the rate to the rest frame. 

We compared the observed SN distribution, binned in redshift, to the expected distribution performing a Maximum Likelihood Estimate to derive the best fit values of the parameters $r(\overline{z})$ and $\alpha$,  independently for SNe~Ia and CC SNe. 

Our candidates come from $V$ and $R$ search programs, with no overlap.
The rates at the reference redshift based on $V$ and $R$ candidates differ by less than 20$\%$ both for type Ia and CC SNe, which is well below the statistical error. Therefore, $V$ and $R$ SN candidates were combined togheter to improve the statistics.

For a {\em standard} extinction, the best fit values and statistical uncertainties for $r(\overline{z})$ and $\alpha$ are:

\begin{itemize}
\item $r^{\rm Ia}(0.30)=0.22^{+0.10}_{-0.08}$ ~$h_{70}^2$ SNu,    ~~~~~$\alpha^{\rm Ia}=4.4^{+3.6}_{-4.0}$ \\ 
\item $ r^{\rm CC}(0.21)=0.82^{+0.31}_{-0.24}$ ~ $h_{70}^2$ SNu,    ~~~~~$\alpha^{\rm CC}=7.5^{+2.8}_{-3.3}$
\end{itemize}

\begin{figure}
\resizebox{\hsize}{!}{\includegraphics{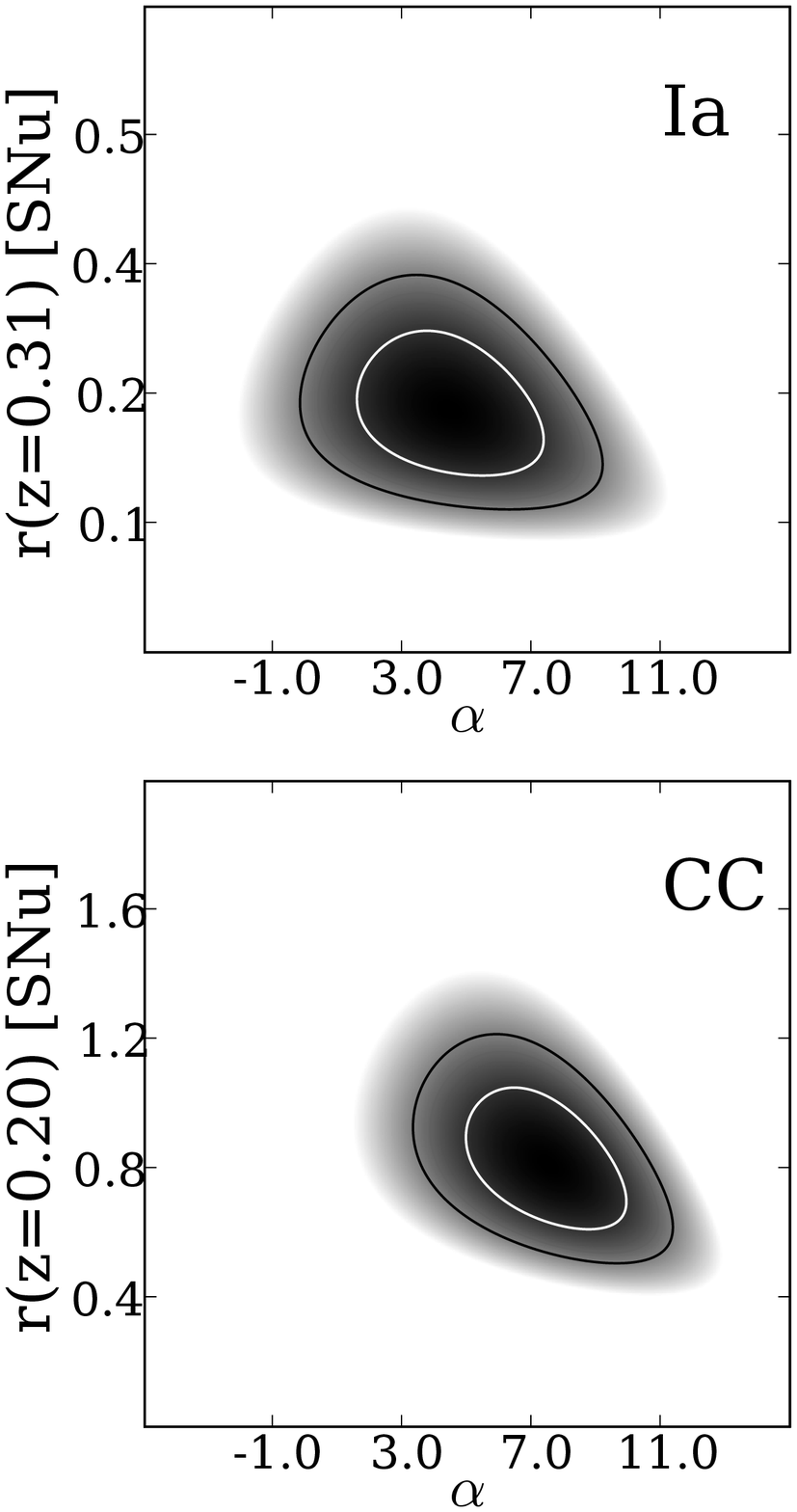}}
\caption{Confidence levels for the parameters describing the redshift evolution of the SN rates, from the maximum-likelihood test.}
\label{livelli}
\end{figure}

\begin{figure}
\resizebox{\hsize}{!}{\includegraphics{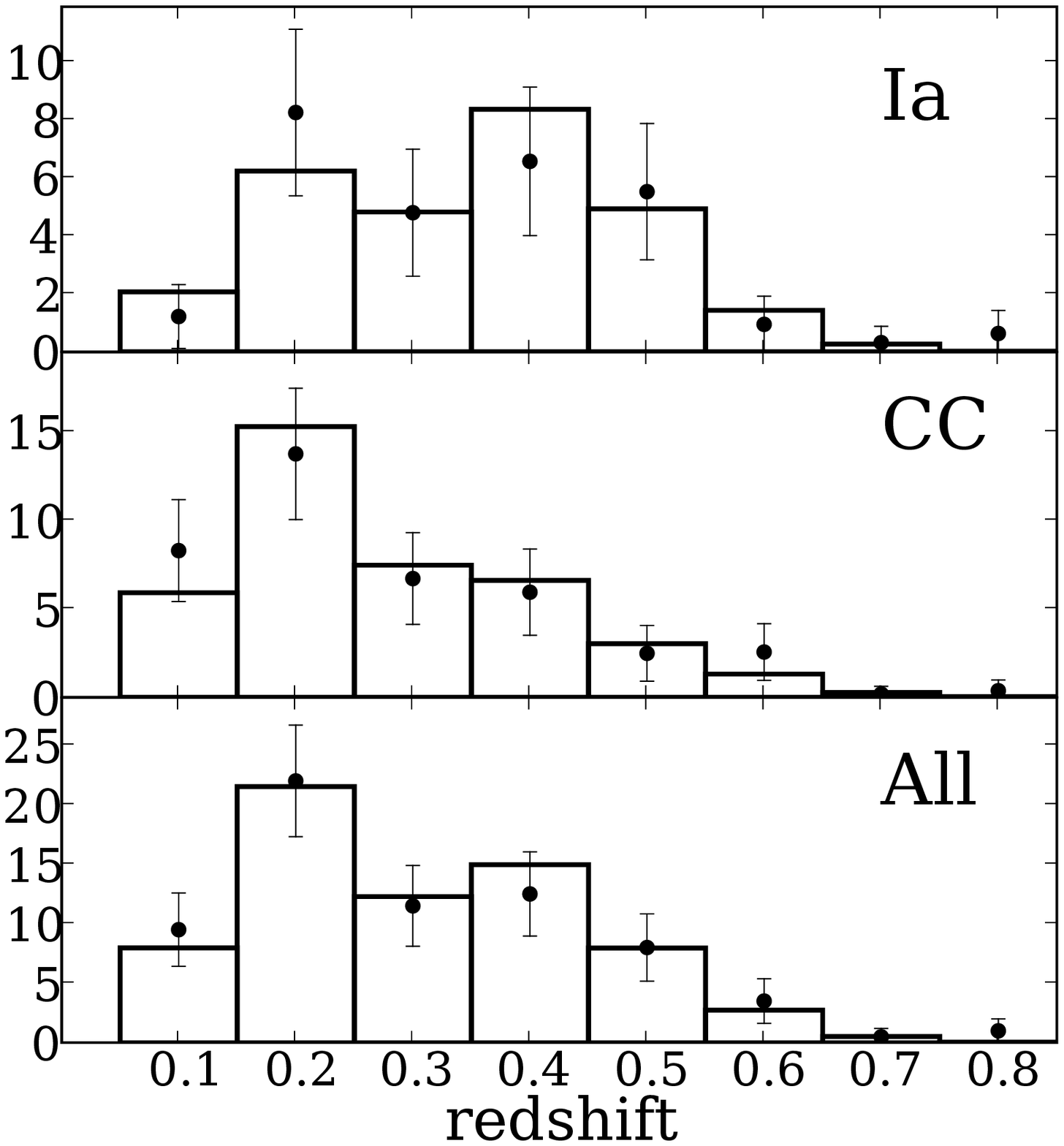}}
\caption{Observed (points with statistical error bars) and expected
(solid line) redshift distribution of SN events.}
\label{fit}
\end{figure}

Fig.~\ref{livelli} shows the 1,2,3 $\sigma$ confidence levels for the two parameters describing the evolution of SN rate, whereas Fig.~\ref{fit} 
shows the best fit of the observed redshift distribution separately for SNe~Ia, CC SNe and then all together. 

The assumption of a particular host galaxy extinction scenario is one of the most important sources of uncertainty for SN rate measurements. We tested the
dependence of the results on this effect by performing the fit under  
extreme assumptions: no extinction and a high extinction scenario,
the latter only for CC SNe.The results are reported in Tab.~\ref{tabres}.
With respect to the standard case, the type Ia and CC SN rates decrese respectively 
by a factor of 1.7 and 2 when no extinction is adopted. 
The effect is smaller for the SN~Ia rate because SNe~Ia occur, on average,
in environments with a smaller amount of dust. If a high extinction correction
is adopted, the CC rate increases by a factor of 2: this can be regarded as
a solid upper limit.

\begin{table}
\centering
\caption{The parameters of the rate evolution estimated for different extinction scenarios.}\label{tabres}
\begin{tabular}{lcccc}
\hline\hline\\
\medskip
SN type     &$\overline{z}$  & Extinction & $r(\overline{z})$  [$h_{70}^2$SNu]   &    $\alpha$ \\
\hline\\
\medskip
Ia         & 0.30    &   none  & 0.13$^{+0.06}_{-0.04}$ & 2.2$^{+3.6}_{-3.8}$\\

\medskip
         &             & standard  &  0.22$^{+0.10}_{-0.08}$ &  4.4$^{+3.6}_{-4.0}$ \\                             
\medskip
CC      & 0.21  & none  & 0.42$^{+0.14}_{-0.12}$ & 5.4$^{+2.8}_{-3.4}$\\   
\medskip
                &       & standard  & 0.82$^{+0.31}_{-0.24}$  &  7.5$^{+2.8}_{-3.3}$ \\     
\medskip
                  &        & high   & 1.66$^{+0.32}_{-0.05}$ & 7.8$^{+2.8}_{-3.2}$ \\
\hline
\end{tabular}
\end{table}

\subsection{Systematic Uncertainties}\label{errors}

When the size of the SN sample is large enough, the statistical uncertainty is 
relatively small, and it becomes important to obtain an accurate estimate of systematic uncertainties. These were evaluated via Monte Carlo simulations. 
We identified the following sources of systematic uncertainties, for which we
specify the distribution adopted for the simulations:

\begin{itemize}
\item  re-distribution of unclassified SNe in SN types (Gaussian distribution with $\sigma$ from the statistical error, as discussed in Sect.~\ref{search}),  
\item AGN contamination factor (Gaussian distribution with an average value of 0.5 and $\sigma=0.25$),
\item photometric redshifts (the number counts in each bin of the galaxy redshift distribution are left to vary by $\pm50\%$ with an uniform distribution), 
\item  detection efficiency (Gaussian distribution with $\sigma=0.1$ mag),
\item SNe absolute magnitude (Gaussian distribution with $\sigma=0.1$ mag),
\item standard extinction correction (uniform distribution with $\pm50\%$ variance with respect to the standard case).
\end{itemize}

The fit described in the previous section was performed for each simulation,
to yield a distribution of solutions for the $r(\overline z)$ 
and $\alpha$. The results are shown in Tab.~\ref{tabunc} for each source of
systematic uncertainty, as derived by varying only one parameter at the
time. The total systematic error has instead been derived by varying all parameters simultaneously.

\begin{table}
\centering
\caption{Summary of systematic uncertainties}\label{tabunc}
\begin{tabular}{lccccc}
\hline\hline
                       & \multicolumn{2}{c}{SNIa} &  &       \multicolumn{2}{c}{SNCC} \\ 
 \cline{2-3}\cline{5-6} 
  Source  item & $\sigma_{r}$ & $\sigma_{\alpha}$& &$\sigma_{r}$ & $\sigma_{\alpha}$ \\
\hline

SN types distribution& +0.08   & +3.4    &   &+0.18     & +3.4\\
 \smallskip
                               &$-0.08$ &$-3.4$ &   &$-0.20$ &$-3.5$\\

AGN fraction                  & +0.02    & +0.2   &   & +0.05   &+0.2 \\
 \smallskip
                              & $-0.01$ &$-0.2$ &  & $-0.05$ & $-0.1$ \\

Photometric redshifts    &+0.04    & +0.7    &  & +0.13 & +0.6\\ 
 \smallskip
                        &$-0.03$ & $-0.7$&   &$-0.11$ & $-0.8$\\

Detection efficiency       & +0.02    & +0.6  &   & +0.21    & +0.6\\
\smallskip
                                      & $-0.03$ & $-0.6$&   & $-0.11$ &$-0.8$\\

Standard extinction       & +0.07    & +1.9  &   & +0.26    & +0.7\\
\smallskip
                                      & $-0.05$ & $-1.4$&   & $-0.22$ &$-0.9$\\

SN absolute mag           & +0.02   & +0.7   &    &+0.06 & +0.4\\
\smallskip
                                      &$-0.02$& $-0.6$ &     &$-0.06$ &$-0.4$\\

 All systematic errors     & +0.16   & +3.3  &     &+0.30 &+3.6\\
 \medskip
                                      &$-0.14$ &$-3.5$ &     &$-0.26$      & $-3.7$\\
Statistical errors             &+0.10    & +3.6   &    &+0.31     & +2.8  \\
                                       &$-0.08$ & $-4.0$&    &$-0.24$  &$-3.2$\\
 \hline
\end{tabular}
\end{table}

Not surprinsingly, the major uncertainty is due to the lack of the spectroscopic classification for a large fraction of the SN candidates. For the CC SN rate also the estimate of the detection efficiency and the dust extinction correction are important sources of uncertainty. 
Even for our relatively small SN sample, the statistical and systematic uncertainties are comparable.
Since, due to a growing number of detected SNe, the statistical uncertainty will decrease in the future, the systematic errors will soon dominate the
overall uncertainty. A special care should then be devoted to reduce the
systematic effects, in particular securing spectroscopic follow up of all the candidates, and obtaining more precise information on the extinction of the galaxy sample.   

\subsection{Comparison with the local SN rates}\label{snrateev}

In order to measure SN rate in a galaxy sample one needs to specify how
the rate scales with a physical parameter proportional to the stellar content of each galaxy. In the early '70 \citet{tamman70,tamman74} and  afterwards \citet{Cappellaro93} showed that the SN rates scale with the galaxy $B$ band luminosity and from then on the SN rates, in the local Universe, have been measured in SNu.
For a direct comparison of the rate at intermediate redshift to the local value
, we also chose to normalize to the galaxy $B$ band luminosity.

The local SN rates measured by collecting data of five photographic SN searches \citep[ and reference therein]{Cappellaro99} are: $r^{Ia}=0.17\pm0.04$ and $r^{CC}=0.41\pm0.17$ $h_{70}^2$SNu at $\overline{z}=0.01$.
As it can be seen the SN~Ia rate in SNu appears constant, within the uncertainties, up to $z=0.3$, whereas the SN~CC rate increases by a factor 2 already at $z=0.21$.
The different evolutionary behaviour of CC and Ia SN rates implies that their ratio increases by a factor of $\sim 2$
from the local Universe to a look-back time of "only" 3 Gyr ($z=0.25$):

\begin{itemize}
\item $(r^{CC}/r^{Ia})_{(0.01)}=2.4\pm1.1$\\
\item $(r^{CC}/r^{Ia})_{(0.25)}=5.6\pm3.5$
\end{itemize}

Considering that, in this same redshift range, the cosmic SFR nearly doubles,
the evolution with redshift of the ratio $r^{CC}/r^{Ia}$ requires that a significant fraction of SN~Ia progenitors has a delay time longer that $2-3$ Gyr (cf. Sect.~\ref{Discussion}).

The interpretation on the evolution of the rate in SNu is not straightforward, as it reflects both the redshift dependence of both the SFH and the $B$-band luminosity.
Indeed, the $B$-band luminosity, with contribution from both old and young stars, evolves with a different slope in comparison with the on-going SFR. 





We acknowledge that the estimate of the SN  rate evolution with  redshift depends on the adopted extinction correction. For instance, the ratio between the rate at z=0.21 and the local rates for CC SNe varies from 1.6 to 2.8, when no or high extinction correction is applied. Anyway, the fact that the CC SN rate increases faster than the SN~Ia rate appears a  robust result.

\subsection{SN rates and galaxy colors}\label{Gcol}

Integrated colours are valuable indicators of the stellar population and SFR in galaxies and, for galaxies at high redshift, the colour information is easier
to derive than the morphological type. Therefore it is interesting to investigate the dependence of the SN rates on the galaxy colours.

In the local Universe, SN rates as a function of galaxy colours were derived by \citet{Cappellaro99} for optical bands and by \citet{Mannucci05} for optical-infrared bands. The SN~Ia rate per unit $B$ luminosity appears almost constant in galaxies with different $U-V$ color, whereas the CC SN rate strongly increases from red to blue galaxies, a trend very similar to that of the SFR. This indicates that a fair fraction of SN~Ia progenitors has long delay times, whereas CC SNe are connected to young massive stars. 

On the other hand, when the local SN rates are normalized to the $K$ band luminosity, or galaxy mass, they show a rapid increase with decreasing $B-K$ galaxy color  for all the SN types \citep{Mannucci05}.  This result indicates that a
sizeable fraction of SNe Ia has short delay times. 
A population of SN~Ia progenitors with short delay time has been proposed also to explain the high SN~Ia rate in radio-loud galaxies \citep{DellaValle},
and \citet{Mannucci06} suggest 
that this fraction ("prompt" component) amounts to $\sim 50 \%$.
In this respect we notice that evolutionary scenarios for SN Ia progenitors do
indeed predict a distribution of the delay times ranging from a few tens of Myr
to the Hubble time, or more, and that such distributions are typically skewed
at the early delay times \citep{Greggio83, Greggio05}. Given the 
diagnostic power of the trend of the SN rates with the colour of the parent 
galaxy we investigate here this trend at intermediate redshift.

In the last decade the extensive study of large samples of galaxies at different redshifts has shown the existence of a bimodal distribution of galaxy colors \citep{Strateva,Weiner,Bell}. In both apparent and rest frame color magnitude diagrams, galaxies tend to separate into a broad blue sequence and a narrow red sequence. While the red sequence is adequately reproduced by passive evolution models, the blue sequence hosts galaxies with different rate of on-going star formation .
Actually, a minor fraction of the red galaxies are also star forming (e.g. edge-on disks or starbusts), their colours being red colours because of high dust extinction. The fraction of star forming galaxies in the red population is $15-20\%$ \citep{Strateva} at low redshift, and  increases to $30\%$ at $z \sim 0.7$ \citep{Bell,Weiner}. The bimodality is not restricted to colour alone, but extends to many other galaxy properties as luminosity, mass and environment, which are all well correlated with colours.

We split up our galaxy sample into blue and red sub-samples, according to the observed $B-V$ color. We took the rest frame color of {\em Sa} CWW template as reference. The local galaxy sample of \citet{Cappellaro99} was divided in the same way. $U-V$ or $B-R$ colours could be more sensitive tracers of the stellar populations,
but $U$-band photometry is not available for our galaxy sample and $R$-band is not available for the local sample. 
 
The SN rates in the blue and red galaxy samples were computed by 
distributing the unclassified SN and SNAGN candidates among type Ia and CC SNe, based on the observed fractions of the spectroscopically confirmed SNe in each of the two galaxy samples (in the red sample SNe~Ia are 67$\%$, whereas they are only 25$\%$ in the blue sample).

The SN rates (in SNu) in the local Universe and at redshift $z=0.25$ for the red and blue galaxy subsamples are listed in Tab.~\ref{tabcolres}.

\begin{table}
\centering
\caption{SN rates as a function of galaxy color [SNu $h^2$]}\label{tabcolres}
\begin{tabular}{lcccc}
\hline\hline
 SN type  &  \multicolumn {2}{c}{$r(z=0.01)$} & \multicolumn {2}{c}{$r(z=0.25)$}  \\
                 & red & blue          & red & blue  \\
\hline
\\
Ia            &$0.20\pm0.04$    &   $0.19\pm0.04$  
               &$0.16^{+0.13}_{-0.18}$ & $0.23 ^{+0.13}_{-0.09}$  \\
\\
CC         &$0.10\pm0.04$   &     $0.99\pm0.15$
              &$0.57^{+0.48}_{-0.30} $   & $1.50^{+0.66}_{-0.48}$ \\
\\
All          &$0.30\pm0.06$             & $1.18\pm0.16 $ 
              &$0.73^{+0.50}_{-0.31}$ & $1.73^{+0.67}_{-0.45}$ \\    
\hline
\end{tabular}
\end{table}

We found very similar trends in the local Universe and at redshift $z=0.25$ both for type Ia and CC SNe. It appears that while the SN~Ia rate is almost costant in galaxies of different colors, the CC SN rate always peaks in blue galaxies. This result is consistent with that of \citet{Sullivan06} who, after comparison with the measurements of \citet{Mannucci05},  found that the SN Ia rate as a function of galaxy colours does not evolve significantly with redshift.

The rapid increase with redshift of the CC SN rate both in blue and red galaxies can be attributed to an increasing proportion of star forming galaxies in the
red sample going from low to intermediate redshift.   

\section{Discussion}\label{Discussion}

In this section we discuss the evolution of the SN rates and investigate on the link between SF and SN rates.
First we verify the consistency of our estimate of the redshift evolution of SN rates with other measurements from the literature  (Sec.~\ref{Complit}).
Then we compare the observed evolution of the CC SN rate with those expected for different SFHs (Sec.~\ref{CompSFR}). Finally, we convolve the SFH of \citet{Hopkins06} with the delay time distribution (DTD) for  different progenitor scenarios, and compare the results to the measurements of the SN~Ia rate evolution (Sec.~\ref{Iamod}). 

\subsection{Comparison with other estimates}\label{Complit}

\begin{figure}
\resizebox{\hsize}{!}{\includegraphics{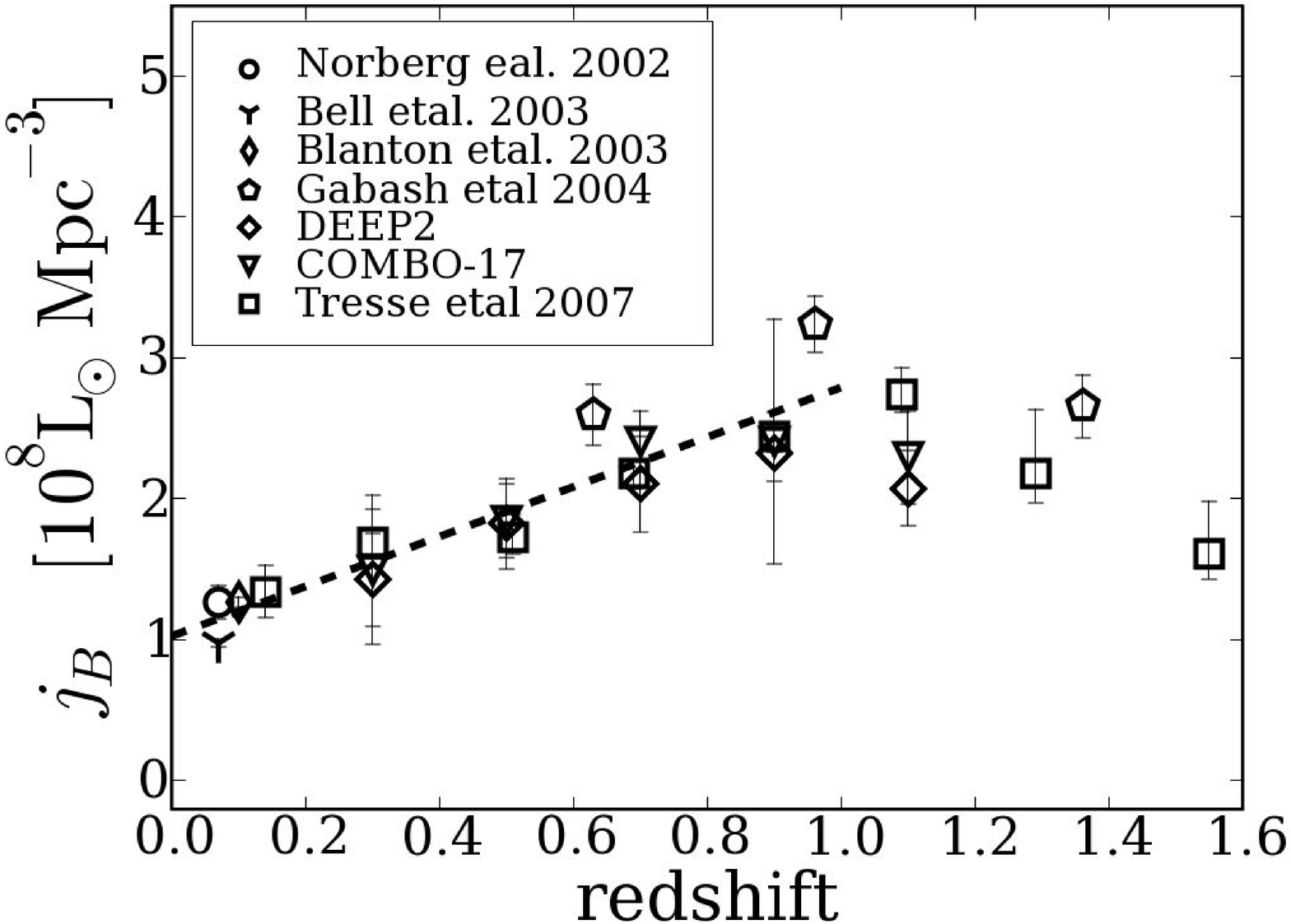}}
\caption{Measurements of the galaxy luminosity density at different redshifts.
The DEEP2 and COMBO-17 data are taken from
Table 2 in \citet{Faber}.
 The line represents the linear least-square fit in the redshift interval $z=0-1$.}
\label{lumdens}
\end{figure}

Published measurements of the SN rates at intermediate and high redshift are
expressed in units of co-moving volume.  To convert our rates from SNu to volumetric rates we assumed that the SN rates are proportional to the galaxy $B$ band luminosity. If this assumption is true, by multiplyng the rates in SNu by the total $B$ band luminosity density in a given volume we derived the total rates in that volume, even if we did not sample the faint end of the galaxy luminosity function.
We accounted for the evolution of the 
$B$ band luminosity density with redshift.
A compilation of recent measurements of the $B$ band luminosity density is plotted as function of redshift in Fig \ref{lumdens}, where we also show 
a linear least-square fit to the data in the range $z=0-1$: $j_B(z) = (1.03+1.76\times z)$ ~$10^8 L_{\odot}^{B}  {\rm Mpc}^{-3}$.  

Multiplying our measurements by the value of $j_B$ at the average 
	    redshifts of the Ia and CC SN samples, the rates per unit of co-moving volume result:  
\begin{itemize}
\item $r^{\rm Ia}(z=0.30) = 0.34^{+0.16 +0.21}_{-0.15 -0.22}$ $10^{-4} h_{70}^3 \mbox{yr}^{-1} \mbox{Mpc}^{-3}$\\
\item$r^{\rm CC}(z=0.21) = 1.15^{+0.43 +0.42}_{-0.33 -0.36}$ $10^{-4} h_{70}^3 \mbox{yr}^{-1} \mbox{Mpc}^{-3}$ 
\end{itemize}
where both statistical and systematic errors are indicated. 
Also local rates are converted into volumetric rates: $r^{\rm Ia}(z=0.01) =0.18\pm0.05$ $10^{-4} h_{70}^3 \mbox{yr}^{-1} \mbox{Mpc}^{-3}$
and $r^{\rm CC}(z=0.01) = 0.43\pm0.17$ $10^{-4} h_{70}^3 \mbox{yr}^{-1} \mbox{Mpc}^{-3}$. 
With respect to the rates in SNu's, the volumetric rates evolve more rapidly 
with redshift, due to the increase of the $B$ band luminosity density. We find an
increase of a factor of $\sim 2$ at $z=0.3$ for SNe~Ia, and a factor of $\sim 3$ at $z=0.21$ for CC SNe.

\begin{figure*}
\resizebox{\hsize}{!}{\includegraphics{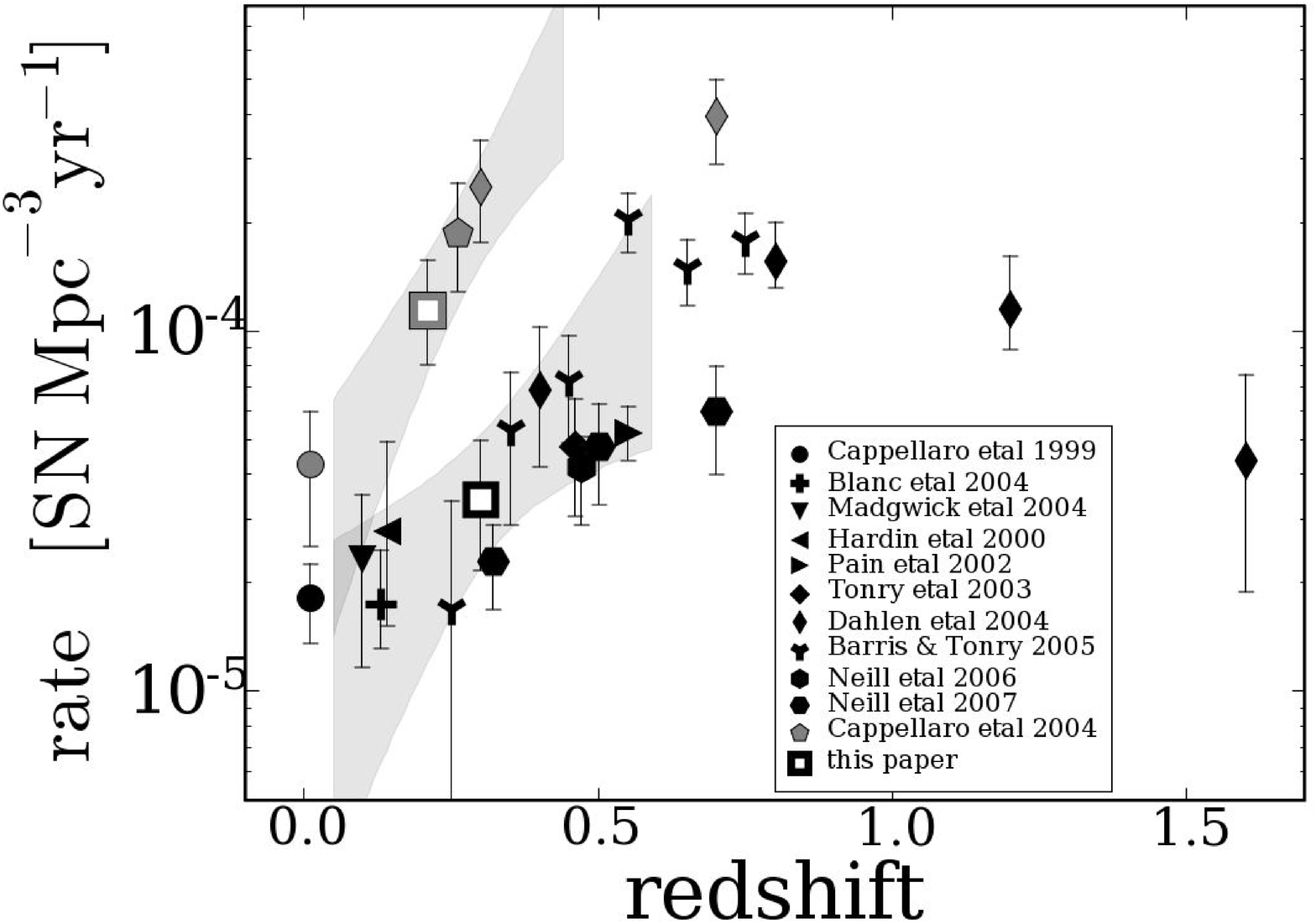}}
 \caption{Observed SN rates as function of redshift from different authors as indicated in the legend. The black (gray) symbols indicate SN~Ia (SN~CC) rate measurements.  The shaded area represents the 1 $\sigma$ confidence level of our rate evolution estimate as deduced from the MLE fit.}
 \label{obsrate}
\end{figure*}

Measurements of Ia and CC SN rate as function of redshift are shown in Fig.~\ref{obsrate}, where those originally given in SNu \citep{Hardin,Blanc,Cappellaro05} were converted into measurements per unit volume as above.
 
As it can be seen, the few measurements of the CC SN rate appear to be fully consistent, while those of the SN~Ia rate show a significant dispersion which increases with redshift, in particular in the range $0.5<z<0.7$ where the values of \citet{Barris} and \citet{Dahlen04} are 2-3 times higher that those of \citet{Pain02} and \citet{Neill07}.
Our estimate of the SN~Ia rate  is consistent  with all other measurements in the redshift range we explored; our result does not help to discriminate
between the steep trend suggested by the \citet{Barris} and \citet{Dahlen04} 
measurements and the slow evolution indicated by the \citet{Neill07} measurement.
The robust indication from the current data appears that the SN~Ia rate per unit volume at redshift 0.3 is a factor of $\sim 2$ higher that in the local Universe, while in the same redshift range the CC SN rate increases by a factor of $\sim 5$.

\subsection{Comparison with the predicted evolution of the CC SN rate}\label{CompSFR}

The stellar evolution theory predicts that all stars more massive than 8-10~M$_\odot$  complete the eso-energetic nuclear burnings, up to the development
of an iron core that cannot be supported by any further nuclear fusion reactions or by electron degenerate pressure. The subsequent collapse of the iron core results into the formation of a compact object, a neutron star or a black hole, accompanied by the high-velocity ejection of a large fraction of the progenitor mass. TypeII SNe originate from the core collapse of stars that, at the time of explosion, still retain their H envelopes, whereas the progenitors of type Ib$/$Ic SNe are thought to be massive stars which have lost their H (and He) envelope \citep{Heger}. Given the short lifetime of their progenitors ($<30$~Myr), there is a simple, direct relation between the CC SN and the current SF rate:

\begin{equation}
r^{\rm CC}(z) = K^{\rm CC} \times \psi(z) 
\end{equation}

where $\psi(z)$ is the SFR and $K^{CC}$ is the number of CC SN progenitors from a 1 $M_\odot$ stellar population:

\begin{equation}
K^{\rm CC} = \frac{\int_{m_l^{\rm CC}}^{m^{\rm CC}_u} \phi(m) dm}{\int_{m_L}^{m_U}m\phi(m) dm}
\end{equation}
where $\phi(m)$ is the IMF, $m_L-m_U$ is the total stellar mass range, and $m^{\rm CC}_l-m^{\rm CC}_u$ is the mass range of CC SN progenitors. 
 
\begin{figure}
\resizebox{\hsize}{!}{\includegraphics{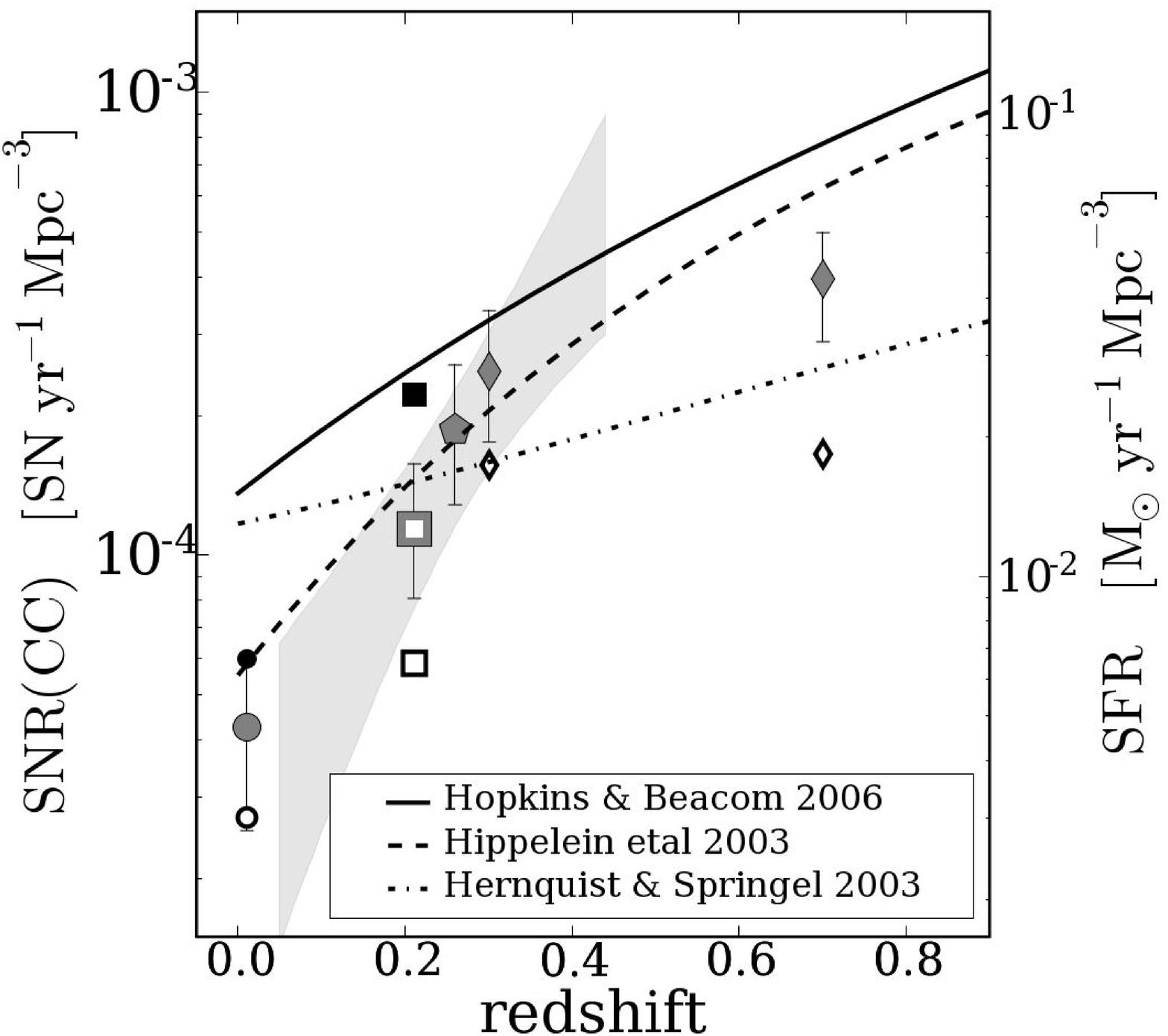}}
 \caption{Comparison between the SN~CC and SF rate evolution. Symbols are as in Fig.~\ref{obsrate} with additional open symbols (measurements not corrected for extinction) and filled black symbols (estimates for the high extinction correction). Lines are selected SFR evolutions from the literature. All SFHs have been scaled to the SalpeterA IMF.}
 \label{ccrate}
\end{figure}

In principle, if accurate measurements of the CC SN and SF rates are available,  it is possible to probe the possible evolution with redshift of either the CC SN progenitor scenarios or the IMF by determining the value of $K^{CC}$.  Here however we assume that $K^{CC}$ does not evolve significantly in the redshift range of interest and compare the observed evolution of the CC SN rate with that predicted by the SFH.

The estimate of the cosmic SFH is based on different SF indicators, depending on redshift range, and requires a suitable parameterization and accurate normalization. 
Since there is a large scatter between the measurements obtained by different SFR indicators (a factor 2-3  at $z=0.3-0.5$ and increasing with redshift) it is  hard to obtain a consistent picture of the SFH \citep{Hopkins06}.  Indeed observations made at different wavelengths, from X-rays to radio, sample different facets of the SF activity and are sensitive to different time scales over which the SFR is averaged. Thus, different assumptions are required  to convert the observed luminosities at the various wavelengths to the SFR, and different systematic uncertainties affect the SFR estimates. 
In particular the significant difference between the SFR inferred from the UV and $H\alpha$ luminosity and that inferred from the far-IR (FIR) luminosity may be related to the effect of dust extinction (expecially for the UV) and/or to the contribution to the light from old stars and AGNs (for the FIR).

To illustrate this point, we select three representative prescriptions for the SFH, namely: 

\begin{itemize}
\item the piecewise linear fit of selected SFR measurements in the range $0<z<6$ by \citet{Hopkins06},
\item the fit to the SFR measurements from the $H\alpha$ emission line, by \citet{Hippelein}, with an exponential increase from $z=0$ to $z=1.2$, 
\item the prescriptions by \citet{Hernquist}, i.e. a double exponential function that peaks at redshift $z\sim 5.5$, obtained from an analytical model and hydro-dynamic simulations.
\end{itemize}

All these SFHs were converted to the same IMF, a modified Salpeter IMF (SalA) with a turnover below 0.5 $M_{\odot}$ and defined in the mass range $m_L=0.1$$M_{\odot}$ to $m_U=120$ $M_{\odot}$   \citep{Baldry}. 
We assumed a  mass range of $8-50 M_{\odot}$ for CC SN progenitors, which gives a scale factor $K^{\rm CC}=0.009$. 

The measurements of the CC SN rate per unit volume and the predicted evolutionary behaviours are shown in Fig~\ref{ccrate}.   The observations confirm the steep increase with redshift expected by the SFH from \citet{Hopkins06} and \citet{Hippelein}.
For a look-back time of 3~Gyr ($z=0.25$) both the SFR and the CC SN rate increase by a factor of $\sim3$ compared with the local values. A flat evolution, as that proposed by \citet{Hernquist}, appears inconsistent with the observed CC SN rates in the overall range of redshift.

With the adopted $K^{\rm CC}$, the level of the 
CC SN rate predicted by the SFH of \citet{Hippelein} and, in general, by the $UV$ and $H\alpha$ based SFHs fits well the data. Instead, for the SFH of \citet{Hopkins06} and in general the SFHs inferred through FIR luminosity, the 
predicted CC SN rate is higher than observed over the entire redshift range \citep[see also][]{Dahlen04,Hopkins06,Mannucci07}.

If we correct the observed CC SN rate according to the high extinction scenario, we obtain an acceptable agreement between the data and the predictions with
\citet{Hopkins06} SFH, as shown in Fig.~\ref{ccrate}.  However, this correction would require an  extremely high dust content in galaxies which is not favored by present measurements.
Indeed, \citet{Mannucci07} derived an estimate of the fraction of SNe which are likely to be missed in optical SN searches because they occur in the nucleus of starburst galaxies or, in general, in regions of very high extinction and found that, at the average redshift of our search, the fraction of missing CC SNe is only $\sim 10\%$, far too small to fill the gap between observed and predicted rates.

Alternatively we may consider the possibility of a narrower range for the CC SN progenitor masses: in particular, a lower limit of $10-12$ $M_{\odot}$ would bring the observed CC SN rates in agreement with those predicted from FIR based SFHs. In this respect we notice that, from a theoretical point of view, there is the possibility that a fraction of stars between 7-8 M$_\odot$ and 10-12 M$_\odot$ avoids the collapse of the core and ends up as ONeMg White Dwarfs \citep{Ritossa,Poelarends}. On the other hand, estimates of the progenitor mass from the detection in pre-explosion (HST)  images has been possible for a few SNe IIP (e.g. SN 2003gd \citep{Hendry}, SN 2005cs \citep{Pastorello}): their absolute magnitudes and colours seem indicate a moderate mass ($8-12$ $M_{\odot}$ )\citep{VanDyk,Smartt,Maund,Li06}. 

Given these controversies, we conclude that in order to constrain the mass range of CC SN progenitors it is necessary to reduce the uncertainties in the cosmic SFH. In addition it is important to apply a consistent dust extinction correction both to SFH and to CC SN rate.

\subsection{Comparison with the predicted evolution of SN Ia rate}\label{Iamod}

According to the standard scenario SNe~Ia originate from the thermonuclear explosion of a Carbon and Oxygen White Dwarf (C-O WD) in a binary system.  In the first phase of the evolution the primary component, a star less massive than $8 M_{\odot}$, evolves into a C-O WD. When the secondary expands and fills its Roche Lobe, two different paths are possible, depending on whether a common envelope forms around the two stars (double degenerate scenario, DD) or not (single degenerate scenario, SD).
In the SD scenario, the WD remains confined within its Roche Lobe, grows in mass until it reaches the Chandrasekhar limit and explodes, while in the DD scenario the binary system evolves into a close double WD system, that merges after orbital shrinking due to the emission of gravitational wave radiation.
 
In both scenarios two basic ingredients are required to model the evolution of the SN~Ia rate: the fraction of the binary systems that end up in a SN~Ia, and 
the distribution of the time elapsed from star formation to explosion (delay-time). In the SD scenario, the delay time is the evolutionary lifetime of the
secondary; in the DD, the gravitational radiation timescale has to be added.
In both cases, the distribution of the delay times depends on the distribution
of the binary parameters \citep{Greggio05}.
In principle, the SD and DD scenarios correspond to different realization probabilities and different shape of the delay time distribution (DTD) functions, hence rather different evolutionary behavior of the SN~Ia rate.
As mentioned previously, the observations indicate that the distribution of the delay times of SN~Ia progenitors is rather wide.

The cosmic SFH is the other critical ingredient that modulates the evolution of the SN~Ia rate. Following \citet{Greggio05} the SN~Ia rate is given by:

\begin{equation}\label{sniaeq}
r^{\rm Ia}(t) = k_{\alpha}A^{\rm Ia}\int_{\tau_{i}}^{min(t,\tau_{x})} f^{\rm Ia}(\tau) \psi(t- \tau)d{\tau}
\end{equation}

where $k_{\alpha}$ is the number of stars per unit mass of the stellar generation, $A^{\rm Ia}$ is the realization probability of the SN~Ia scenario (the number fraction of stars from each stellar generation that end up as SN~Ia), $f^{\rm Ia}(\tau)$ is the distribution function of the delay times and $\psi(t-\tau)$ is the star formation rate at the epoch $t-\tau$. The integration is extended over all values of the delay time $\tau$ in the range $\tau_{i}$ and $min(t,\tau_{x})$, with $\tau_{i}$ and $\tau_{x}$ being the minimum and maximum possible delay times for a given progenitor scenario. Here we assumed that both $k_{\alpha}$ and $A^{\rm Ia}$ do not vary with cosmic time.

A detailed analysis of the predicted evolution of the SN~Ia rate for different SFHs and DTDs is presented elsewhere \citep{Forster,Blanc07}. Here we consider only one SFH, the piecewise interpolation of \citet{Hopkins06},  since this is conveniently defined also at high redshift, and limit our analysis to few DTDs representative of different approaches to model the SN~Ia rate evolution: three models from \citet{Greggio05}, and two different parametrizations \citep{Mannucci06,Strolger} designed to address some specific observational constraints, regardless the  correspondence to a specific progenitor scenario.

Specifically, among the \citet{Greggio05} models we select one SD and two DD-models, one of the ``close'' and the other of the ``wide'' variety, the latter being an
example of a relatively flat DTD. This choice is 
meant to represent the full range of plausible DTDs. The minimum delay
time for the SD model is the nuclear lifetime of the most massive secondary stars in the SN Ia progenitor's system, i.e. $8 M_{\odot}$. In principle, for the DD model the minimum delay time could be appreciably larger than this because of the additional gravitational waves radiation delay. In practice, also for the DDs the minimum delay is
of a few $10^{7}$ yrs. The maximum delay time is quite sensitive to the model for the SN~Ia progenitors.
The reader should refer to \citet{Greggio05} for a more detailed description of these models.

The DTD parametrization proposed by \citet{Mannucci06} is the sum of two distinct functions: a Gaussian centered at 5$\times10^{7}$ yr, representative of a ``prompt'' progenitor population which traces the more recent SFR, and an exponentially declining function with characteristic time of 3 Gyr, a ``tardy'' progenitor population proportional to the total stellar mass. This DTD was introduced to explain, at the same time, the dependence of the SN~Ia rate per unit mass on the galaxy morphological type, the cosmic evolution of the SN~Ia rate and, in particular, the high SN~Ia rate observed in radio-loud galaxies.

Finally, \citet{Strolger} showed that the best fit of the apparent decline of SN~Ia rate at $z> 1$ is achieved for a DTD with a Gaussian distribution centered at about 3~Gyr. Nevertheless this DTD fails to reproduce the dependence of the SN rate on galaxy colours which is observed in the local Universe \citep{Mannucci06}.

\begin{figure}
 \resizebox{\hsize}{!}{\includegraphics{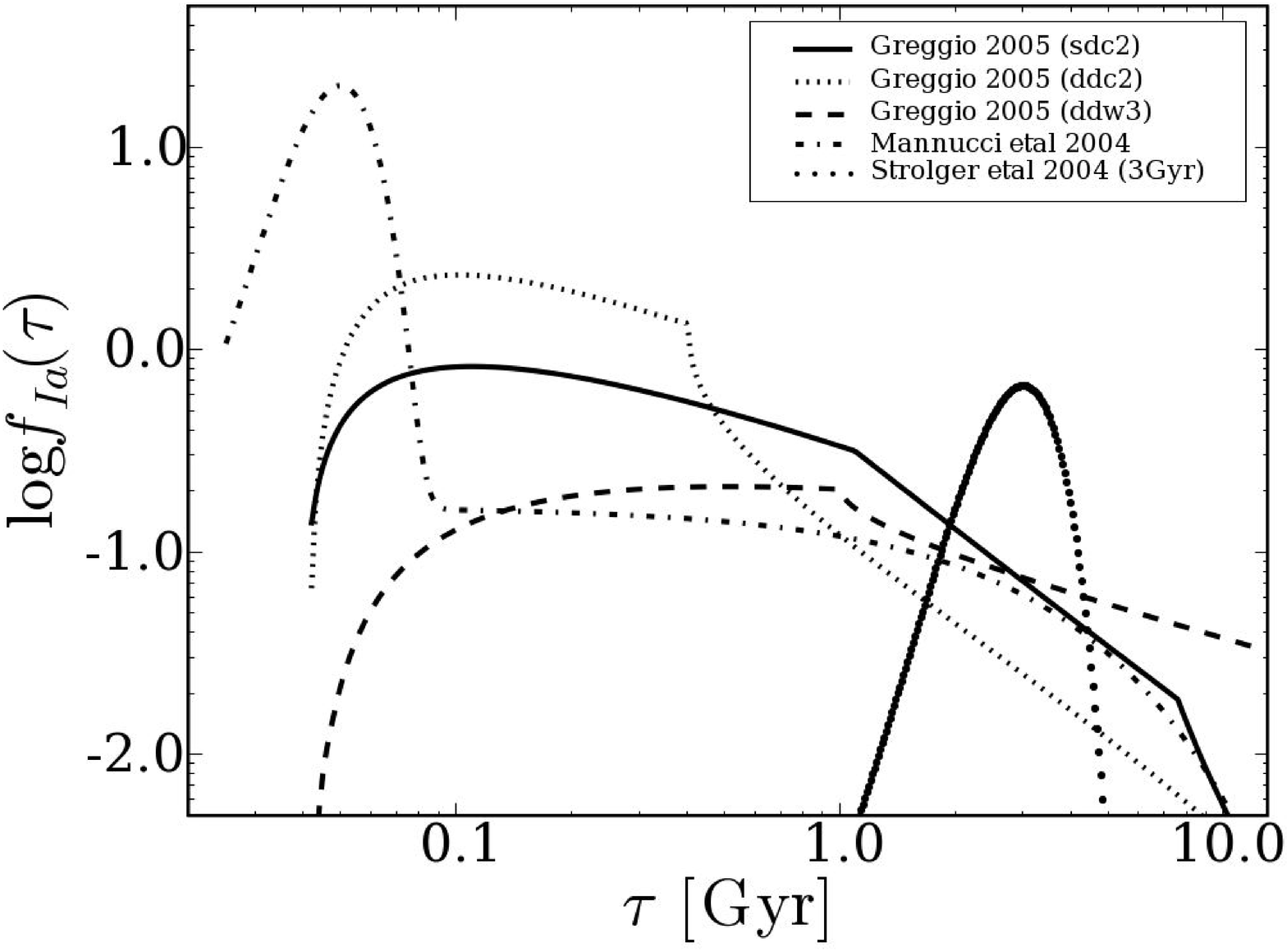}}
 \caption{Delay time distributions as derived by \citet{Greggio05} for SD and DD models, \citet{Mannucci06} and \citet{Strolger}.}
 \label{delay}
\end{figure}

\begin{figure}
 \resizebox{\hsize}{!}{\includegraphics{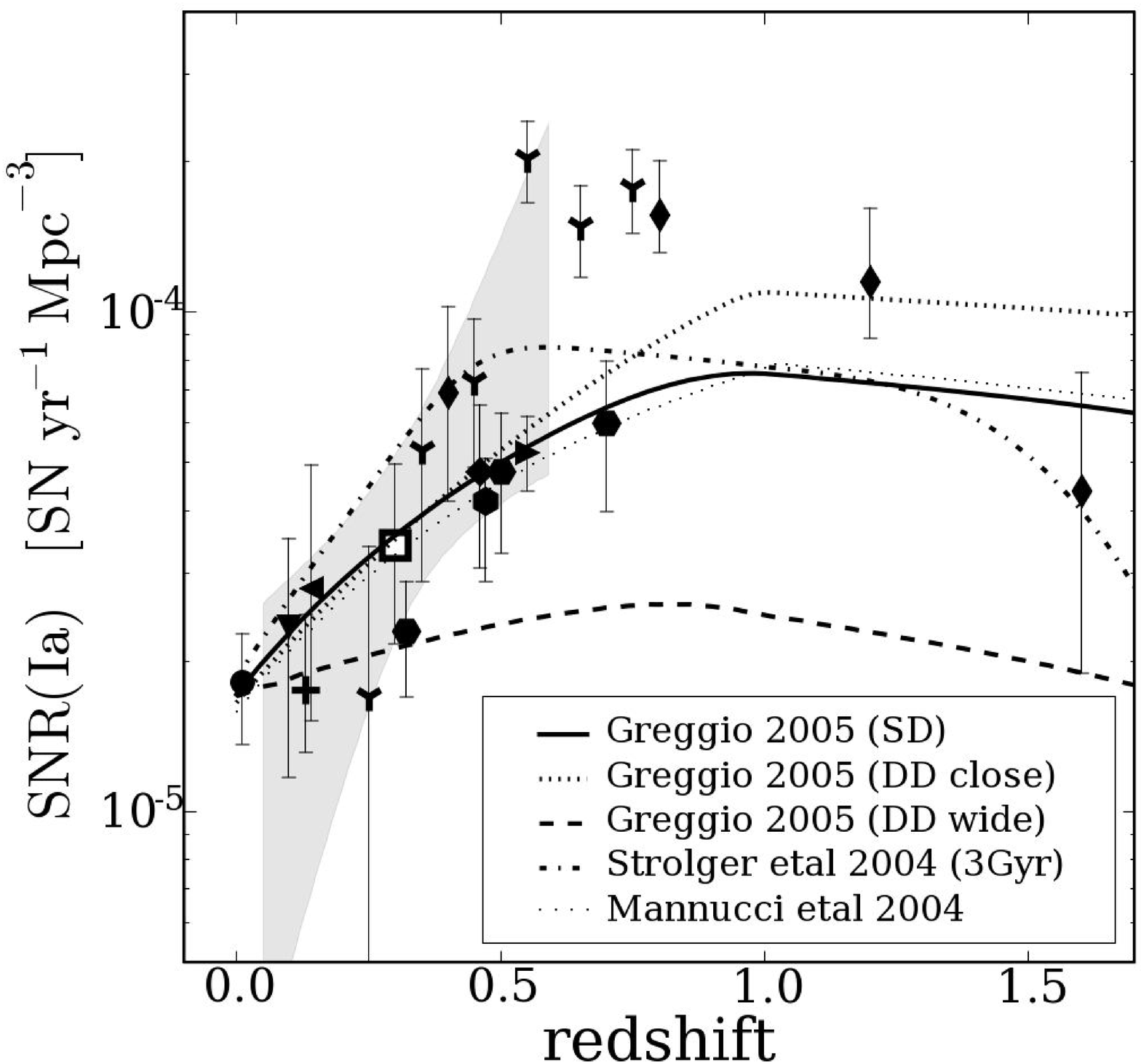}}
 \caption{SN~Ia rate measurements fitted with different DTD functions and the SFH by \citet{Hopkins06}. Symbols for measurements are as in Fig.~\ref{obsrate}.}
 \label{rateia}
\end{figure}

The selected DTD functions are plotted in Fig.~\ref{delay} while the predicted evolutionary behaviours of the SN~Ia rate are compared with all published measurements in Fig.~\ref{rateia}.  In all cases, the
 value of $k_{\alpha}A^{\rm Ia}$ was fixed to match the value of the local rate; depending on the model it ranges between 3.4-7.6$\times 10^{-4}$. This normalization implies that, for the adopted SalA IMF, and assuming a mass range for the
progenitors of $3-8$ $M_\odot$,  the probability that a star with suitable mass becomes a SN~Ia, is $\sim$ 0.01-0.03.

The models obtained with the different DTDs are all consistent with the observations with the exception of the "wide" DD model, whose redshift evolution is definitely too flat. On the other hand none of the DTD functions, with the adopted SFH, is able to reproduce at the same time the very rapid increase from redshift 0 to 0.5 suggested by some measurements \citep{Barris,Dahlen04} and the decline at redshift $>1$ \citep{Dahlen04}.
We note that a new measurement of \citet{Poznanski} suggests that the SN~Ia rate decline at high redshift may be not as steep as estimated by \citet{Dahlen04}.

Given the current uncertainties of both SN~Ia rate and SFH it is difficult to discriminate between the different DTD functions and hence between the different SN~Ia progenitor models. To improve on this point, more measurements of SN~Ia rate at high redshifts are required  to better trace the rate evolution. At the
same time measurements in star forming and in passive evolving galaxies in a wide redshift range can provide important evidence about the SN~Ia progenitor models.
In addition, it is essential to estimate the cosmic SFH more accurately because the position of the peak of the SFH was found to be the crucial parameter for the recovered delay time \citep{Forster}.

\section{Summary}\label{Summary}

In this paper we have presented our SN search (STRESS) carried out with ESO telescopes and aimed at measuring the rate for different SN types at intermediate redshift. 

Our approach, that consists in counting the SNe observed in a well defined galaxy sample, allows us to investigate the dependence of the rates on galaxy colours, and to perform a direct comparison with measurements in the local Universe, which are obtained with the same approach.

For the selection and the characterization of the galaxy sample we used multi-band observations and the photometric redshift technique. After 16 observing runs, 12 in $V$, 4 in $R$ band, we discovered a few hundred variable sources from which we selected 86 SN candidates: 25 spectroscopically confirmed SNe (9 SNe~Ia and 16 CC SNe), 33 SN candidates and 28 SNAGN candidates.

The control time of our galaxy sample has been evaluated through a detailed analysis of the SN detection efficiency of our search, with a number of MonteCarlo simulations. In addition, we have improved the handling of the correction for dust extinction by elaborating realistic modelling. The rates are presented 
for two scenarios, one stardard and one which, maximizing the effect, yield 
reasonable upper limits.

Since our SN sample spans a wide redshift range ($0.05<z<0.6$) we have obtained an observational constraint on the redshift evolution of the rates, described as a power law with two parameters: $r(\overline{z})$, the rate at the mean redshift of the search $\overline{z}$, and $\alpha$, an evolution index.
For SNIa and CC SNe we get: 

\begin{description}
\item  SN~Ia:  $r(\overline{z}=0.3)= 0.22^{+0.10 +0.16}_{-0.08 -0.14}$ ~$h_{70}^2$ SNu,                       ~$\alpha=4.4^{+3.6 +3.3}_{-4.0 -3.5}$  

\item CC SN: $r(\overline{z}=0.2)= 0.82 ^{+0.31 +0.30}_{-0.24 -0.26}$  ~$h_{70}^2$ SNu,  ~$\alpha=7.5^{+2.8 +3.6}_{-3.2 -3.7}$ 
\end{description}

where both statistical and systematic uncertainties are reported.

Our results indicate that, compared to the 
local value, the CC SN rate per unit $B$ band luminosity increases by a factor of $\sim 2$ already at $z\sim 0.2$, whereas the SN~Ia rate is almost constant up to redshift $z\sim 0.3$. 
The dependence of the SN rates per unit $B$ band luminosity on the galaxy colour is the same as observed in the local Universe: the SN~Ia rate seems to be almost constant from red to blue galaxies, whereas the CC SN rate seems to peak in the blue galaxies.
Therefore, on the one hand, the observed evolution with redshift of the ratio $r^{CC}/r^{Ia}$ requires a significant fraction of SN~Ia progenitors with long delay times; on the other hand, the observed trend with galaxy colour requires a short delay time for a fraction of SN~Ia progenitors.  

Systematic uncertainties in our SN rate measurements have been investigated by considering the influence of the different sources. In particular we have analysed the important role played by the AGN contamination of SN candidate sample and the host galaxy extinction.

After accounting for the evolution with redshift of the blue luminosity density, we have compared our estimates with all other measurements at intermediate redshift available in the literature and found that they are consistent, within the relatively large errors.

Finally we have exploited the link between SFH and SN rates to predict the evolutionary behaviour of the SN rates and compare it with the path indicated by observations.

The predicted evolution of the CC SN rate with redshift has been computed assuming three representative SFHs and the mass range 8-50 $M_\odot$ for CC SN progenitors. The comparison with the observed evolution confirms the steep increase with redshift indicated by most recent SFH estimates. Specifically, we found a good agreement with the predictions from the SFH inferred through H$\alpha$ luminosity \citep{Hippelein}, while the SFHs from FIR luminosity  \citep{Hopkins06} overestimate the CC SN rate, unless a higher extinction correction and/or a narrower range for the progenitor masses is adopted.
This result illustrates how interesting clues can be obtained by 
comparing the SN~CC rate to other SFR tracers, in the same galaxy sample, 
to verify the reliability of the techniques used to derive SFR estimates and 
the adequacy of the dust extinction corrections.

The cosmic evolution of the SN~Ia rate has been estimated by convolving the SFH of \citet{Hopkins06} with various formulations of DTD related to different SN~Ia progenitor models. All DTDs appear to predict a SN~Ia rate evolution consistent with the observations, with the exception of the "wide" DD model, which 
appears too flat. At the same time none of the explored DTD functions, at least with the adopted SFH and $k_{\alpha}A^{\rm Ia}$ factor are able to reproduce simultaneously both the rapid increase from redshift 0 to 0.5 and the decline at redshift $>1$ suggested by some measurements. 

With the current data of the rate evolution, it is difficult to discriminate between different DTDs and then between different SN~Ia progenitor models. 
Measurements of the SN~Ia rate in star forming and in passive evolving galaxies in a wide range of redshifts can provide more significant evidence about SN~Ia progenitors. 

An extensive survey to search all SN types in a well characterized sample of galaxies is of the highest priority to probe the link between the SN rates and the SFR in different contexts. This will yield important constraints on SN progenitor scenarios from a detailed analysis of SN rates as a function of the various properties of galaxies.

\begin{acknowledgements}
MTB acknowledges Amedeo Tornamb\'{e} for useful discussions and the European Southern Observatory for hospitality during the development of this work. 
This work has been supported by the Ministero
dell'Istruzione, dell'Universit\'{a} e della Ricerca (MIUR-PRIN 2004-029938,2006-022731) and by the National Science Foundation under Grant No. PHY05-51164.
G.P acknowledges support by the Proyecto FONDECYT 3070034.
This research has made use of the NASA/IPAC
Extragalactic Database (NED) which is operated by the Jet Propulsion
Laboratory, California Institute of Technology, under contract with
the National Aeronautics and Space Administration. 

\end{acknowledgements}

\Online

\end{document}